A review of Fe-chalcogenide superconductors: the simplest Fe-based superconductor

Yoshikazu Mizuguchi[1,2,3], Yoshihiko Takano[1,2,3]
1. National Institute for Materials Science, 1-2-1 Sengen, Tsukuba, 305-0047, Japan
2. Japan Science and Technology Agency-Transformative Research-Project on Iron-Pnictides (JST-TRIP), 1-2-1 Sengen, Tsukuba, 305-0047, Japan
3. Graduate School of Pure and Applied Sciences, University of Tsukuba, 1-1-1 Tennodai, Tsukuba, 305-8571, Japan

Abstract
Here we summarize the physical properties of the newly discovered Fe-chalcogenide superconductors. The Fe-chalcogenide superconductors attract us as the simplest Fe-based superconductors. Furthermore, Fe chalcogenides show a huge pressure effect on their superconducting properties. The origin of the high transition temperature was discussed with both the change in crystal structure and magnetism. The progress on the thin-film and superconducting-wire fabrications are also described.

1. Introduction: superconductivity in FeSe

Since the discovery of superconductivity in LaFeAsO$_{1-x}$F$_x$, FeAs compounds which have a structure similar to LaFeAsO have been confirmed to show superconductivity at the comparably high transition temperature $T_c$.[1-6] They commonly contain anti-PbO-type FeAs layers as the superconducting layers in its crystal structure. In July 2008, Hsu et al. reported the superconductivity at 8 K in anti-PbO-type FeSe.[7] Figure 1 shows the temperature dependence of resistivity, and the insets indicate the resistivity under the magnetic fields and the estimated upper critical field $\mu_0H_{c2}$. The high $\mu_0H_{c2}$, which was one of the common natures of the Fe-based superconductivity, was observed also for FeSe. FeSe is composed of only the FeSe layers with the anti-PbO-type structure (space group: $P4/nmm$) as drawn in Fig. 2. The crystal structure of FeSe is the simplest among Fe-based superconductors. Furthermore, the theoretical study indicated the similarity in the electronic states between Fe chalcogenides (FeS, FeSe and FeTe) and the FeAs-based superconductors.[8] The contributions of Fe-3d electrons near the Fermi level $E_F$ and the morphology of the Fermi surface exhibited similarities to the FeAs-based superconductors as shown in Fig. 3. Because of these natures common to the FeAs-based superconductors, the FeSe superconductor has attracted a lot of researchers as the key material to elucidate the mechanism of Fe-based superconductivity.

Here we summarize the superconducting properties of Fe chalcogenides, mainly based on the experimental results. At first, the physical properties of Fe chalcogenides at ambient pressure are summarized in chapter 2 with the several phase diagrams. In chapter 3, the pressure effects are discussed with the correlation between superconductivity and magnetism. The relationship between superconducting properties and crystal structure are also discussed with anion height dependence of $T_c$. The superconducting gaps observed by the photoemission spectroscopy (PES) and the scanning tunneling spectroscopy (STS) are described in chapter 4. At the end of this article, the reports of thin film and superconducting wire fabrications are described in chapter 5.

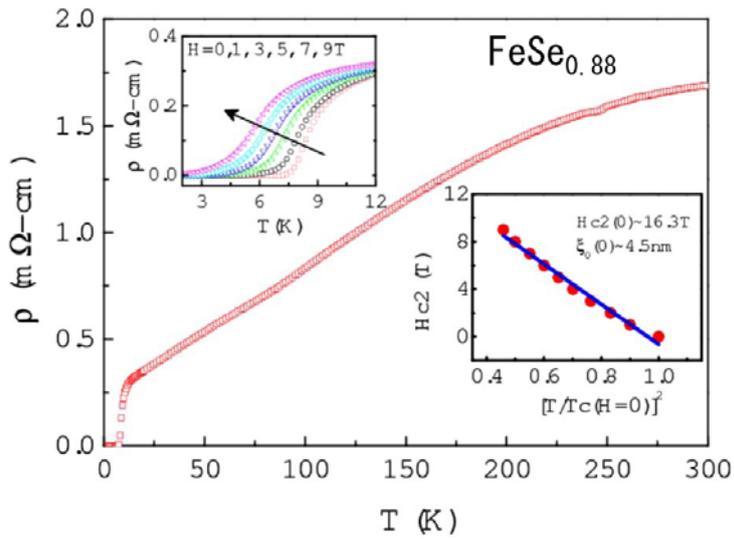

Fig. 1. Temperature dependence of resistivity for FeSe$_{0.88}$. The insets display the temperature dependence of resistivity under magnetic fields and the estimated $H_{c2}$. [Figure reprinted from F. C. Hsu et al., Proc. Natl. Acad. Sci. 105, 14262 (2008). Copyright 2008 by The National Academy of Sciences of the USA.]

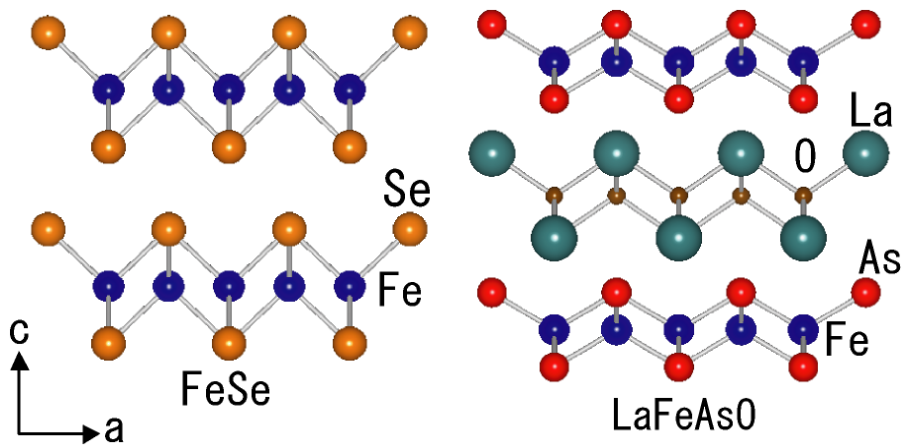

Fig. 2. Crystal structure of FeSe and LaFeAsO. The figures were drawn using VESTA.[9]

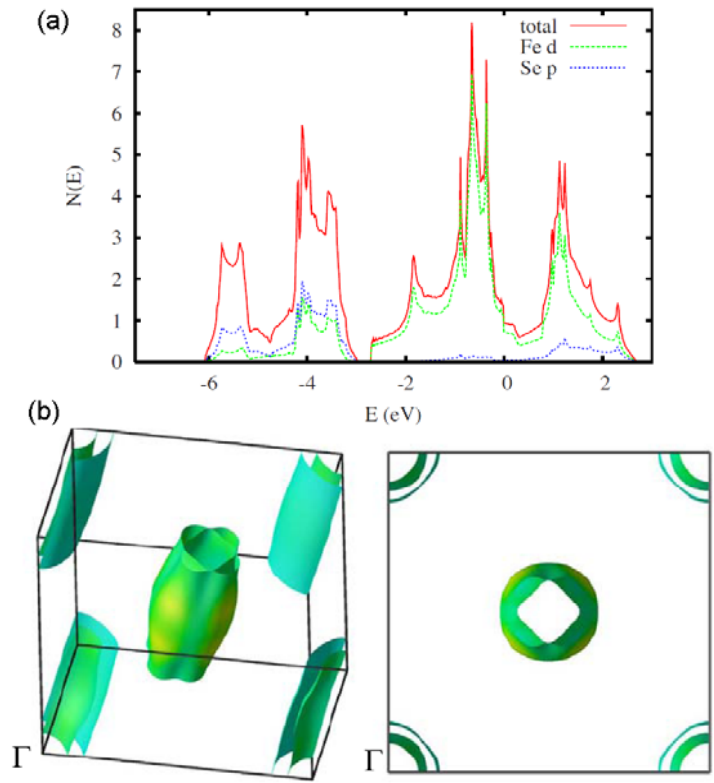

Fig. 3. (a)Electronic density of state (DOS) and projection onto the linearized augmented plane-wave (LAPW) Fe and chalcogen spheres for FeSe. (b)Local-density approximation (LDA) Fermi surface of FeSe from non-spin-polarized calculations with the LDA relaxed $X$ heights.

[Figure reprinted from A. Subedi et al., Phys. Rev. B 78, 134514 (2008). Copyright 2008 by The American Physical Society.]

## 2. Phase diagram of Fe chalcogenide

### 2-1. Variation of Fe chalcogenide

Fe chalcogenides form several types of crystal structures, according to the elemental composition, synthesis process and synthesis conditions of temperature or pressure. To date, three anti-PbO-type compounds, FeS, FeSe and FeTe,[10-13] have been confirmed. Among them, FeTe is the most stable phase, and the single phase can easily be obtained by a conventional solid-state reaction method. With decreasing ionic radius of chalcogen, the PbO structure tends to be unstable. Although FeSe forms with the solid-state reaction, the sample synthesized at high temperatures contains the NiAs-type (hexagonal) FeSe phase. To obtain the single phase of PbO-type FeSe, low temperature annealing around 300 - 400 ºC, which transforms the NiAs-type phase to the PbO-type phase, is required.[14] Figure 4 shows the sintering-temperature dependence of the superconducting transition in the magnetization measurement. Sample (a) was reacted at 1100 ºC and then annealed at 400 ºC for 200 h. Sample (b) and (c) were reacted at 1100 ºC and 680 ºC, respectively, and these compounds contain the NiAs phase. The superconducting transition for sample (a) was the sharpest, and the complete shielding was observed, indicating that both the high-temperature reaction and low-temperature annealing were required to obtain the high-quality FeSe sample. Furthermore, PbO-type FeS is much unstable and cannot be obtained by the solid state reaction. It forms only by a chemical process in an aqueous solution.[11]

### 2-2. Phase diagram of $Fe_{1+d}Te_{1-x}Se_x$

We discuss the physical properties of the end-member compounds FeSe and FeTe. Although FeSe and FeTe have similar crystal structure, their physical properties are much different. Figure 5 shows the temperature dependences of resistivity for FeSe and FeTe. FeSe shows a metallic behavior and undergoes superconducting transition at $T_c^{onset}$ = 13 K. On the other hand, FeTe exhibits the antiferromagnetic ordering around 70 K where the anomaly appears in the resistivity-temperature curve, and does not show superconductivity.[15-17]

Figure 6 shows the $^{57}$Fe-mössbauer spectra for FeSe from room temperature down to 4.2 K.[18] The major features were fitted by a single paramagnetic doublet (Isomer shift $IS$ = 0.538(6) mm/s, quadrupole splitting $QS$ = 0.268(10) mm/s at 4.2 K), indicating an absence of the magnetic ordering above 4.2 K for superconducting FeSe. On the other hand, FeTe shows an antiferromagnetic ordering below 70 K. Figure 7 shows the mössbauer spectra for FeTe from room temperature down to 4.2 K. Below the structural and magnetic transition temperature, a clear magnetic sextet (hyperfine field $H_{hf}$ = 103.4(11) kOe at 4.2 K) corresponding to the antiferromagnetic ordering was observed. The comparably low internal magnetic field implies the low-spin state of Fe.

Crystal structure analysis at low temperatures for FeSe indicated an existence of the structural transition from tetragonal to orthorhombic around 70-90 K as displayed in Fig. 8.[19,20] For $Fe_{1.068}Te$, the structural transition from tetragonal to monoclinic was confirmed by the neutron diffraction as shown in Fig. 9.[21] Also the magnetic spin structure was determined by the neutron diffraction as shown in Fig 10. The spin structure of FeTe is different from that of the FeAs-based parent compounds. The properties of FeTe depend on the content of excess Fe at the interlayer site as described as the Fe(2) site in Fig. 11(a). The tetragonal-monoclinic structural transition and the antiferromagnetic ordering with a magnetic wave vector $q$ of (1/2 0 1/2) were confirmed for $Fe_{1.076}Te$. For $Fe_{1.141}Te$, which had the excess-Fe concentration higher than that of $Fe_{1.076}Te$, the tetragonal-orthorhombic structural transition and an incommensurate magnetic wave vector $q$ of ($\pm\delta$ 0 1/2) were reported as shown in Fig. 11.[22] Here we focus on the $Fe_{1+d}Te_{1-x}Se_x$ system with low $d$, because superconductivity tends to be suppressed with higher content of excess Fe.[23]

To establish a phase diagram of the mixed phase $Fe_{1+d}Te_{1-x}Se_x$, we summarize their physical properties. Figure 12 shows the temperature dependence of resistivity for $FeTe_{1-x}Se_x$.[17] While the samples were almost single phase, the superconducting transitions were broad for the primitive polycrystalline samples, implying an existence of the local phase separation. In fact, as shown in Fig. 13, there is a miscible region, at which the phase separation occurs, around $x = 0.7 \sim 0.95$.[15-17] The crystal structure at low temperatures were determined using the single-phase samples with $x = 0$-0.57. With increasing Te concentration, the tetragonal-orthorhombic structural transition observed in FeSe was suppressed. For $FeTe_{0.43}Se_{0.57}$, the structural transition temperature $T_s$ was determined to be 40 K by the synchrotron x-ray diffraction as described in Fig. 14.[24] Li et al. reported that the tetragonal-orthorhombic transition disappeared for $FeTe_{0.507}Se_{0.493}$, and the short-range antiferromagnetic fluctuations grew up with further increase of Te content.[21] For $FeTe_{0.743}Se_{0.257}$, the peak corresponding to the short-range antiferromagnetic appeared at fluctuations wave vector $Q$ = 0.938 Å$^{-1}$, which was slightly less value than the $Q$ value of 0.974 Å$^{-1}$ at the commensurate position as shown in Fig. 15.[21] Prassides et al. also reported similar observation of the short-range antiferromagnetic peak at Q = 0.935 Å$^{-1}$.[25] For $Fe_{1.06}Te_{0.87}Se_{0.13}$, the long-range antiferromagnetic ordering and the tetragonal-monoclinic structural transition were observed.[25]

The actual superconducting transition temperature was determined by the magnetic susceptibility measurement for the high-quality single crystals. For the Te-rich region ($x < 0.5$) the large single crystals could be grown, and the sharp superconducting transition with $T_c$ = 14 K was obtained for FeTe$_{0.5}$Se$_{0.5}$ crystal synthesized by the Bridgman method as shown in Fig. 16.[26] Recently, Taen et al. reported the annealing effect on the superconducting properties of the single crystals grown by the melting method.[27] As reported in FeSe, the annealing around 400 ºC was required to achieve bulk superconductivity for the melted crystals. We have synthesized the FeTe$_{1-x}$Se$_x$ crystals by the melting method with 400 ºC annealing for $x$ = 0.13, 0.25 and 0.35. Figure 17 shows the temperature dependence of magnetic susceptibility at the zero-field cooling (ZFC) and field cooling (FC) for the plate-like single crystals. The actual compositions were determined by the energy dispersive x-ray spectroscopy (EDX).

By summarizing these experimental results, we established the phase diagram of Fe$_{1+d}$Te$_{1-x}$Se$_x$ with low excess-Fe concentration as shown in Fig. 18. The tetragonal-orthorhombic structural transition observed in FeSe is suppressed with increasing Te concentration. The highest $T_c$ appears at the tetragonal phase near $x$ = 0.5. With further increase of the Te content, the $T_c$ decreases and the antiferromagnetic ordering accompanying the tetragonal-monoclinic distortion grew up, and the bulk superconductivity disappears.

2-3. Phase diagram of FeTe$_{1-x}$S$_x$

As the partial Se substitution for Te suppresses the antiferromagnetic ordering and induced superconductivity for FeTe, the S substitution for Te also suppresses magnetic ordering and induces superconductivity.[28] Figure 19 shows the temperature dependence of resistivity for Fe$_{1.08}$Te, FeTe$_{0.9}$S$_{0.1}$, FeTe$_{0.8}$S$_{0.2}$ synthesized using the melting method, where the compositions are the starting nominal values. The antiferromagnetic ordering observed in FeTe was suppressed by the S substitution, and superconductivity was observed at $T_c^{onset}$ ~ 10 K. Figure 20 shows the temperature dependence of magnetic susceptibility for FeTe$_{1-x}$S$_x$. While zero resistivity was observed in the superconducting samples, the superconducting volume fraction estimated from the magnetic susceptibility was less than 20 % for FeTe$_{0.8}$S$_{0.2}$, implying the inhomogeneity of the sample and/or the insufficiency of S doping. The solid-state reaction method allowed synthesizing the almost single phase, but the solid-state reacted FeTe$_{1-x}$S$_x$ showed only filamentary superconductivity. The difficulty in synthesis of FeTe$_{1-x}$S$_x$ would be due to the solubility limit of S for the Te site probably arising from the difference of the ionic radius between S and Te. Figure 21 is a phase diagram of FeTe$_{1-x}$S$_x$. The S substitution for Te suppresses magnetic ordering in FeTe and induces superconductivity, as in FeTe$_{1-x}$Se$_x$.

Moisture-induced superconductivity was observed for the solid-state-reacted FeTe$_{0.8}$S$_{0.2}$ sample, which showed only a broad onset of the superconducting transition. Surprisingly, the superconducting properties dramatically improved by just exposing the sample to the air.[29] Figure 22 and 23 show the temperature dependence of resistivity and magnetic susceptibility, respectively, for solid-state-reacted FeTe$_{0.8}$S$_{0.2}$ with several air-exposure days. The $T_c^{zero}$ appeared after a few days and reached 7.2 K after 110 days. Also the superconducting volume fraction estimated from the magnetic susceptibility measurement was enhanced from 0 to 48.5 %. By measuring the susceptibility for the samples kept in vacuum, water, O$_2$ gas and N$_2$ gas for several days, we concluded that the moisture in the air induced bulk superconductivity in solid-state-reacted FeTe$_{0.8}$S$_{0.2}$ because only the sample kept in water showed the diamagnetic signal corresponding to superconductivity. Elucidation of origin of this phenomenon will provide us key information to understand the relationship between the appearance of superconductivity and the magnetism or the crystal structure for Fe chalcogenides.

Recently, we found that the oxygen annealing at ~200 ºC also induced superconductivity for solid-state-reacted FeTe$_{0.8}$S$_{0.2}$.[30] Furthermore, the oxygen-annealed FeTe$_{0.8}$S$_{0.2}$ showed a sharp superconducting transition and a high shielding volume fraction of almost 100 % as shown in Fig. 24. The temperature dependence of resistivity also changed by the oxygen annealing, and a broad hump was observed around 100 K, as observed in optimally doped FeTe$_{1-x}$Se$_x$. By using the oxygen annealing process, we will be able to obtain FeTe$_{1-x}$S$_x$ series with the several doping level, and establish an accurate phase diagram.

2-3. Phase diagram of FeSe$_{1-x}$S$_x$

The S substitution for Te suppresses the magnetic ordering in FeTe and induced superconductivity. How does the S substitution for the Se site affect the superconducting properties of FeSe? Figure 25 shows the temperature dependence of resistivity for FeSe$_{1-x}$S$_x$, where the compositions are the starting nominal values. With S substitution, the tetragonal-orthorhombic transition observed in FeSe was suppressed for FeSe$_{0.9}$S$_{0.1}$, and the $T_c$ slightly increased up to 20 %.[17,31] However, the $T_c$ was suppressed above 20 % substitution, and zero-resistivity state was not observed for FeSe$_{0.6}$S$_{0.4}$. The phase diagram of FeSe$_{1-x}$S$_x$ was established as shown in Fig. 26. The S substitution seems to be disadvantageous for superconductivity of FeSe, compared to the Te substitution for the Se site.

## 2-4. Overview of superconductivity in Fe chalcogenides

Here we summarize the superconducting properties of Fe chalcogenides. The highest $T_c$ appears in optimally doped $FeTe_{1-x}Se_x$, which has the tetragonal structure down to low temperatures. While the higher $T_c$ appears in the tetragonal phase, the orthorhombic phase of $FeTe_{1-x}Se_x$ ($0.5 \leq x \leq 1$) also can be superconducting. However, in monoclinic phase of FeTe, the long-range antiferromagnetic ordering appears and superconductivity is not observed. These results suggest that the higher $T_c$ appears in the tetragonal phase close to magnetically ordered FeTe. The theoretical study predicted the higher stability of magnetism and possibility of higher $T_c$ for FeTe than for FeSe and FeS.[8] The magnetic fluctuation should be the key to understand the variation of $T_c$ in this system, and it is likely to be much important to discuss the dramatic pressure effects of Fe chalcogenides, which will be described in the next chapter.

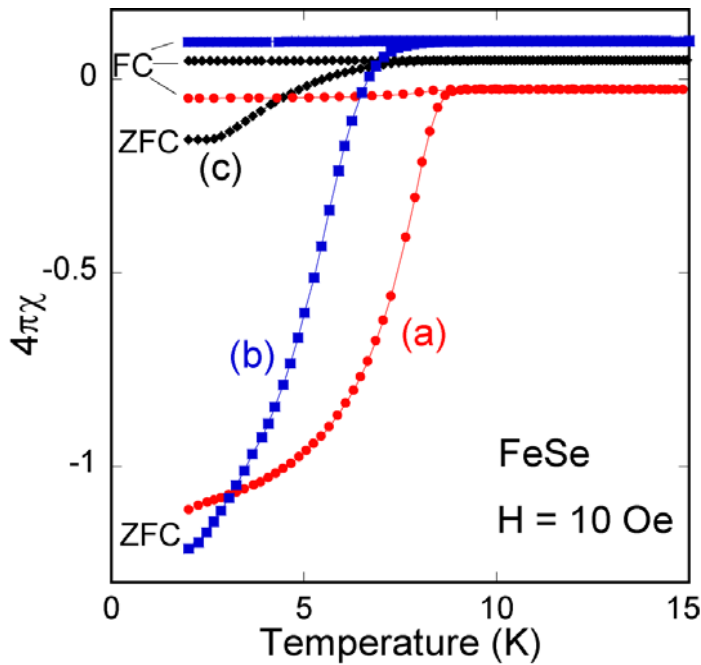

Fig. 4. Temperature dependence of magnetic susceptibility for the FeSe sample synthesized by three heating processes. (a) 1100 ºC + 400 ºC for 200 h. (b) 1100 ºC. (c) 680 ºC.

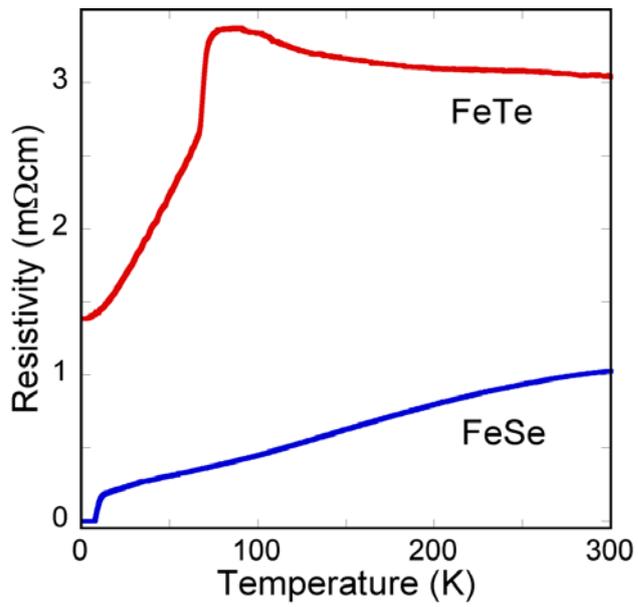

Fig. 5. Temperature dependence of resistivity for FeSe and FeTe. FeSe shows metallic behavior and undergoes superconducting transition. In contrast, FeTe exhibits antiferromagnetic ordering around 70 K and does not show superconductivity.

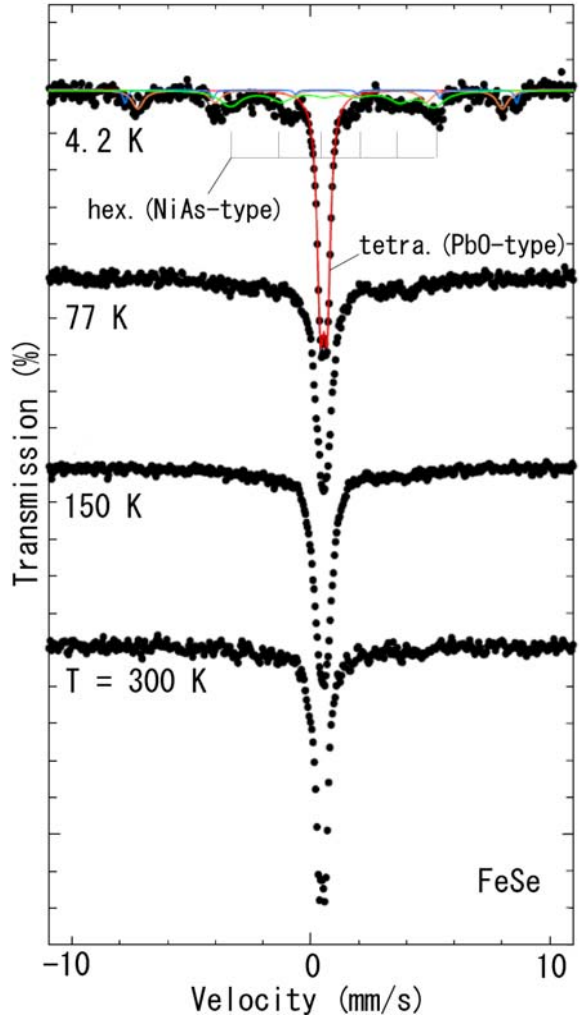

Fig. 6. $^{57}$Fe Mössbauer spectra of FeSe at 300, 150, 77 and 4.2 K. There is no sign of magnetic ordering for the major phase of tetragonal FeSe as indicated by the red fitting curve. For the spectrum at 4.2 K, small magnetic sextet with $H_{hf}$ = 264.7(80) kOe, which is fitted with green fitting curve, corresponds to the signal of the minor-phase hexagonal FeSe. The minor magnetic sextets indicated by blue or orange curves correspond to the signals of Fe oxides with $Fe^{3+}$ in high-spin states.

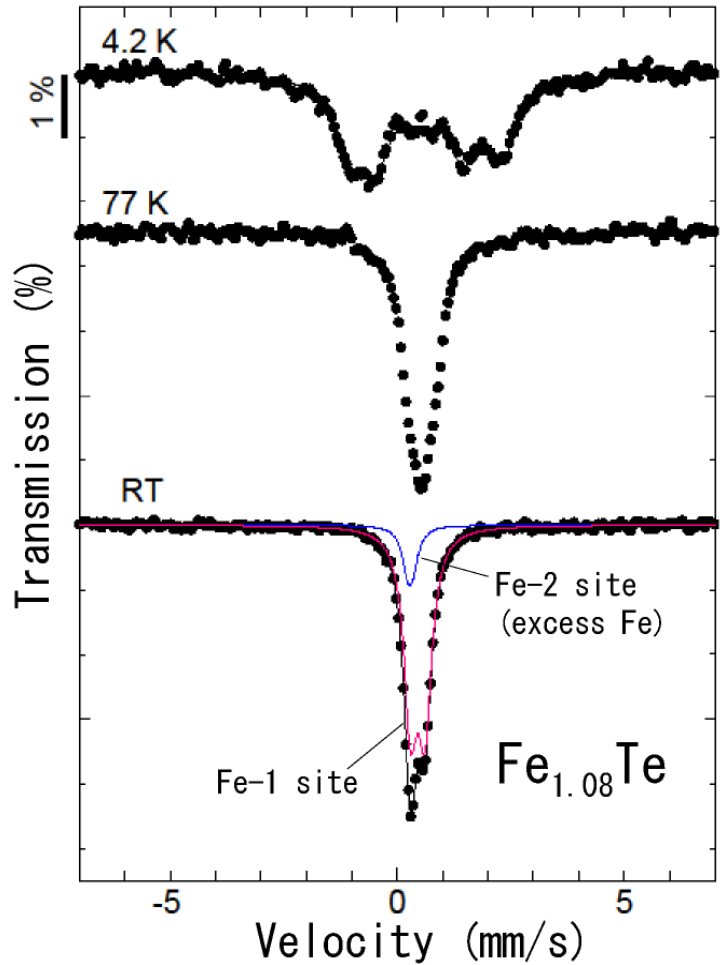

Fig. 7. $^{57}$Fe Mössbauer spectra of Fe$_{1.08}$Te at room temperature, 77 and 4.2 K. The spectrum at room temperature is fitted by two types of doublets, which are attributed to the two Fe sites. One site is the Fe in the FeTe layer (Fe-1 site), and the other is Fe which exists at the interlayer site (Fe-2 site). At 4.2 K, the clear magnetic sextet corresponding to the magnetic ordering around 70 K was observed.

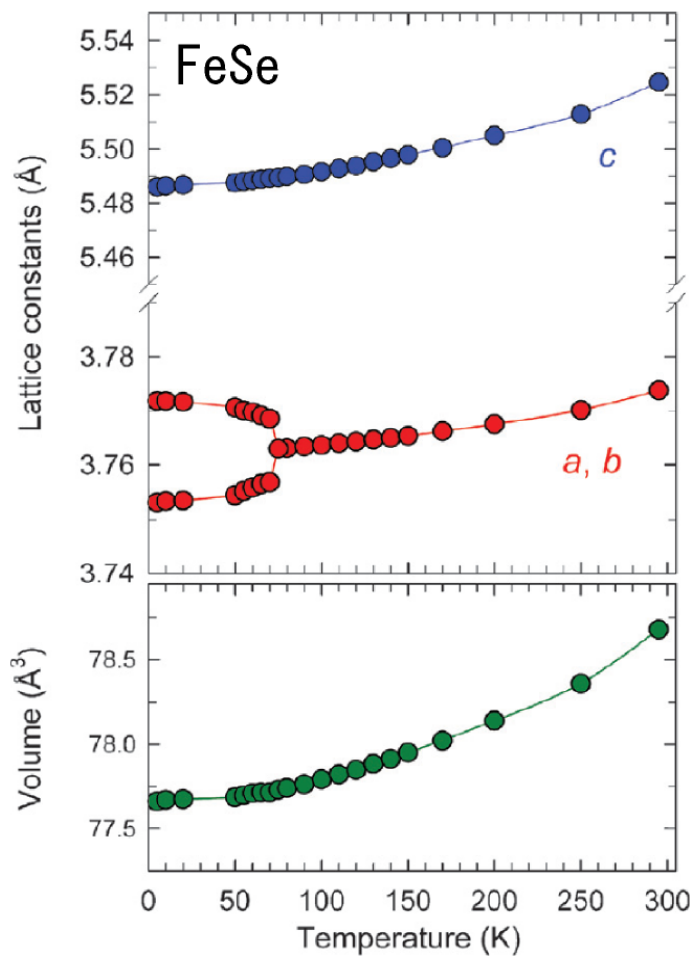

Fig. 8. Temperature dependence of lattice constants determined by the synchrotron x-ray diffraction or the neutron diffraction. The structural transition from tetragonal to orthorhombic was observed around 70 K. The *a* and *b* lattice constants are divided by √2 at temperatures below the tetragonalto-orthorhombic phase transition.
[Figure reprinted from S. Margadonna et al., Chem. Commun. 2008, 5607. Copyright 2008 by The Royal Society of Chemistry.]

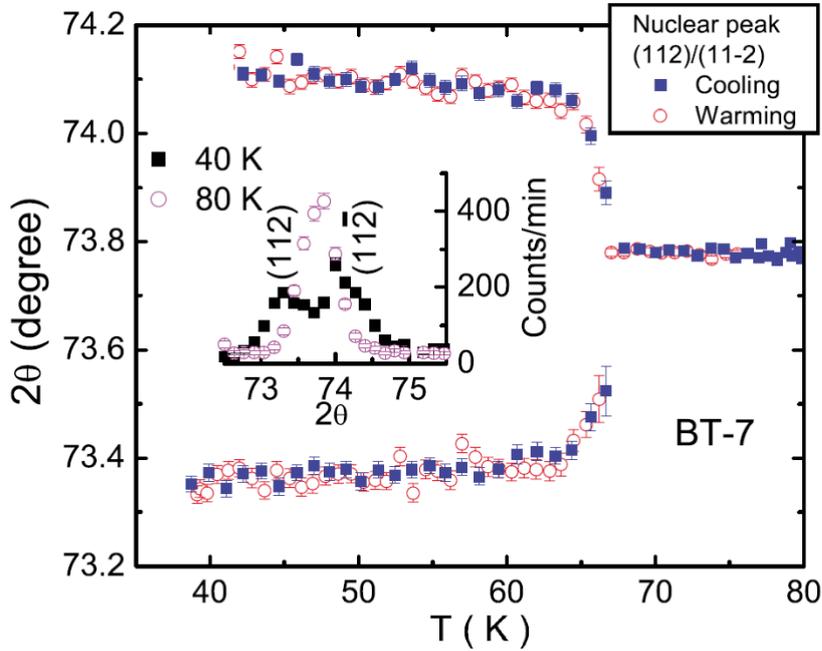

Fig. 9. Splitting of the (1,1,2) and (1,1,−2) nuclear peaks with decreasing temperature due to the tetragonal-monoclinic lattice distortion.
[Figure reprinted from S. Li et al., Phys. Rev. B 79, 054503 (2009). Copyright 2009 by The American Physical Society.]

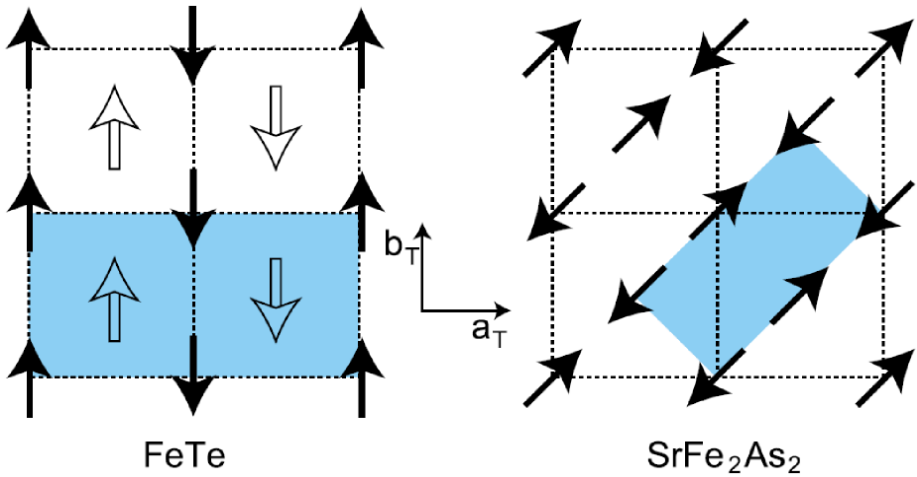

Fig. 10. Schematic in-plane spin structure of $Fe_{1.068}Te$ and FeAs-based $SrFe_2As_2$. The solid arrows and hollow arrows represent two sublattices of spins, which can be either parallel or anti-parallel. The shaded area indicates the magnetic unit cell.
[Figure reprinted from S. Li et al., Phys. Rev. B 79, 054503 (2009). Copyright 2009 by The American Physical Society.]

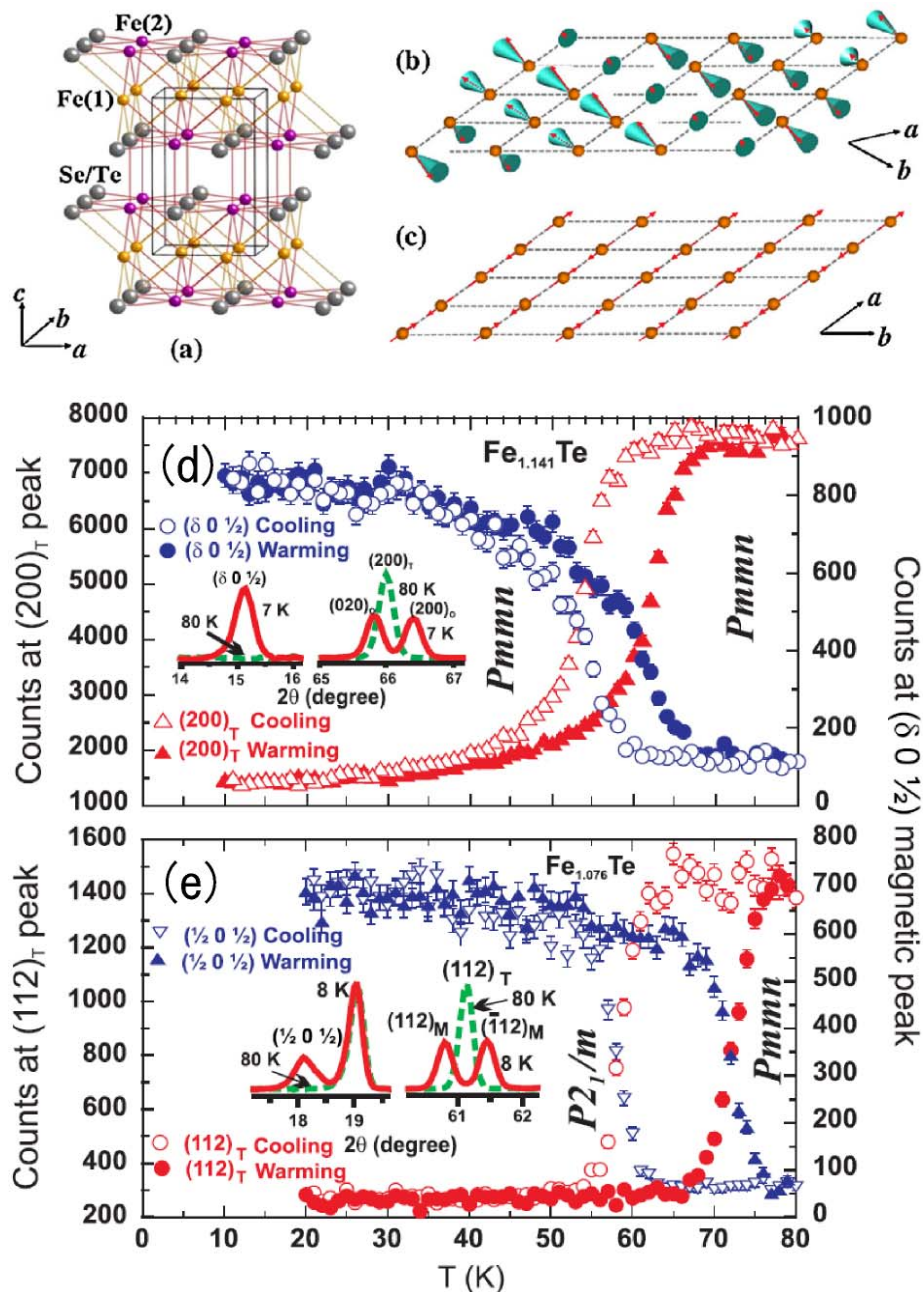

Fig. 11. (a) Crystal structure of Fe(Te,Se). Magnetic structures of (b) FeTe and (c) BaFe$_2$As$_2$ are shown in the primitive Fe square lattice for comparison. Note that the basal square lattice of the PbO unit cell in (a) is $\sqrt{2} \times \sqrt{2}$ superlattice of that in (b). (d),(e) The magnetic Bragg peak ($\delta$, 0, 1/2) (blue symbols) and the splitting of the structural peak (200) or (112) of the tetragonal phase (red symbols) show the thermal hysteresis in the first-order transition.
[Figure reprinted from W. Bao et al., Phys. Rev. Lett. 102, 247001 (2009). Copyright 2009 by The American Physical Society.]

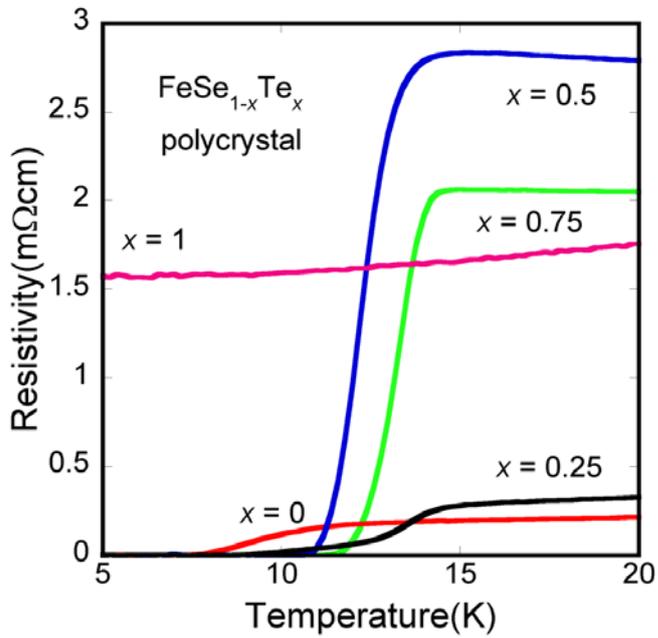

Fig. 12. Temperature dependence of resistivity for FeTe$_{1-x}$Se$_x$.

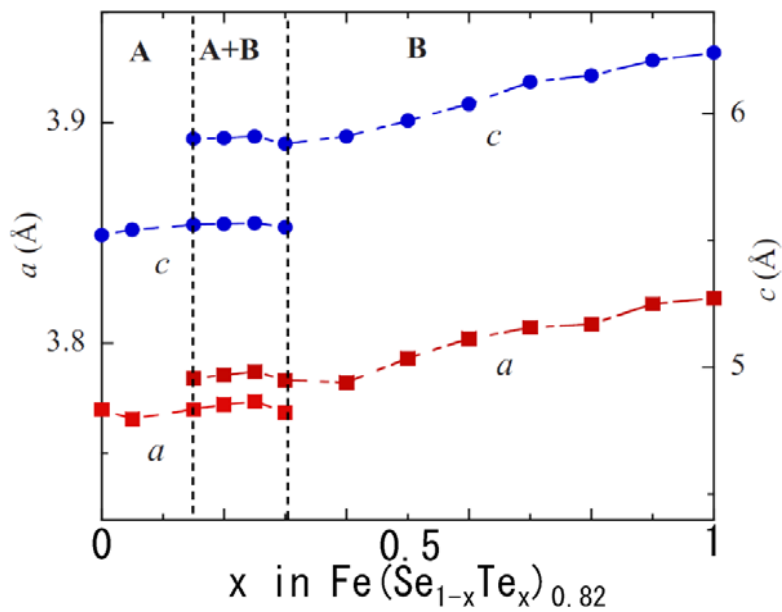

Fig. 13. Composition dependence of lattice constants $a$ and $c$. The miscible region exists with $x$ = 0.05-0.3 in Fe(Se$_{1-x}$Te$_x$)$_{0.82}$.
[Figures reprinted from M. H. Fang et al., Phys. Rev. B 78, 224503 (2008). Copyright 2009 by The American Physical Society.]

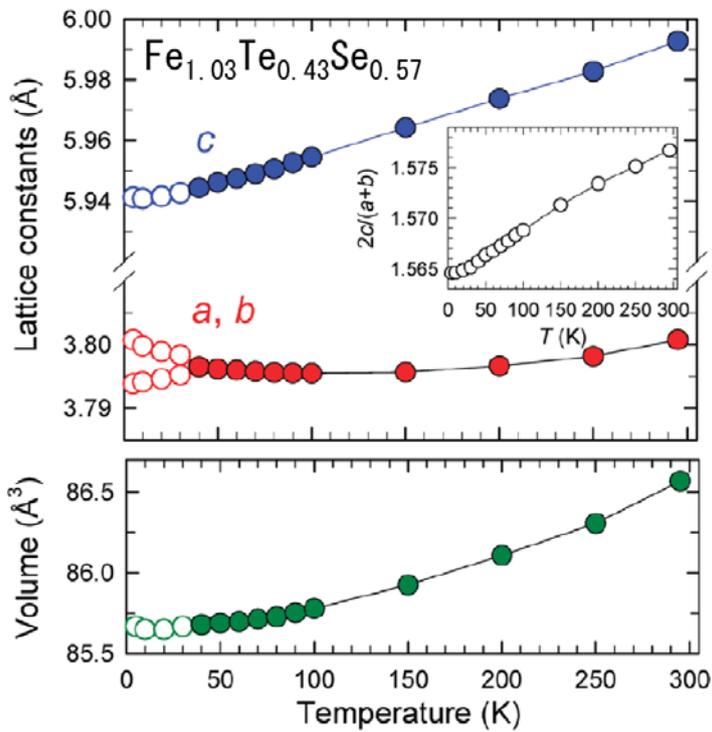

Fig. 14. Temperature evolution of the lattice constants (top) and the unit cell volume, $V$ (bottom), in $Fe_{1.03}Te_{0.43}Se_{0.57}$. The inset shows the temperature dependence of the $2c/(a+b)$ ratio. $a$ and $b$ are divided by $\sqrt{2}$ and $V$ by 2 at temperatures below the tetragonal-to-orthorhombic phase transition at 40 K.
[Figure reprinted from N. C. Gresty et al., J. Am. Chem. Soc. 131, 16944 (2009). Copyright 2009 by American Chemical Society.]

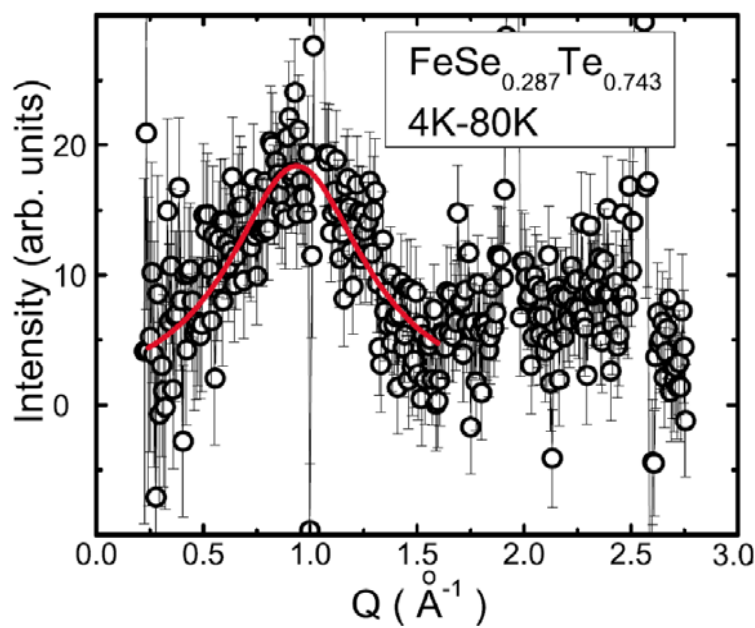

Fig.15. Short-range antiferromagnetic fluctuations at $Q$ = 0.938 Å$^{-1}$ in FeTe$_{0.743}$Se$_{0.287}$.
[Figure reprinted from S. Li et al., Phys. Rev. B 79, 054503 (2009). Copyright 2009 by The American Physical Society.]

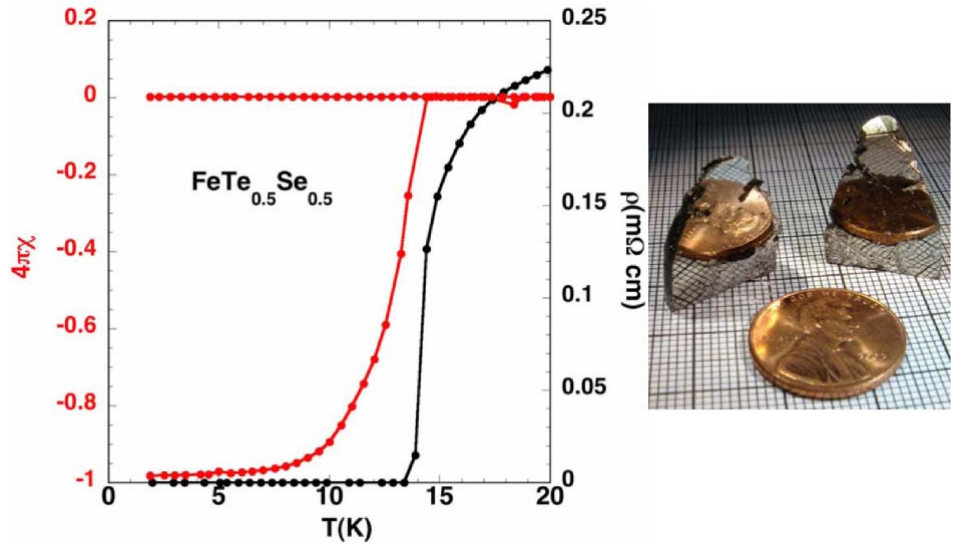

Fig. 16. Left figure shows the superconducting transition of FeTe$_{0.5}$Se$_{0.5}$ single crystal grown by the Bridgman method. Right picture displays the cleaved surface of FeTe$_{1-x}$Se$_x$ crystal.
[Figures reprinted from B. C. Sales et al., Phys. Rev. B 79, 094521 (2009). Copyright 2009 by The American Physical Society.]

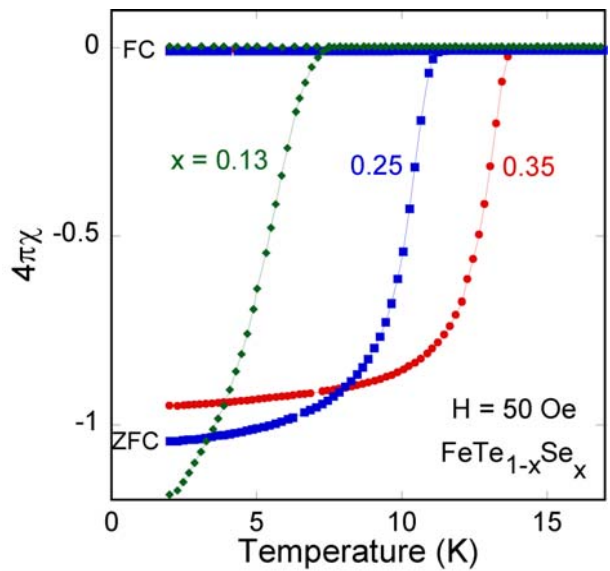

Fig. 17. Temperature dependence of magnetic susceptibility $\chi$ for the FeTe$_{1-x}$Se$_x$ crystals grown by the melting method.

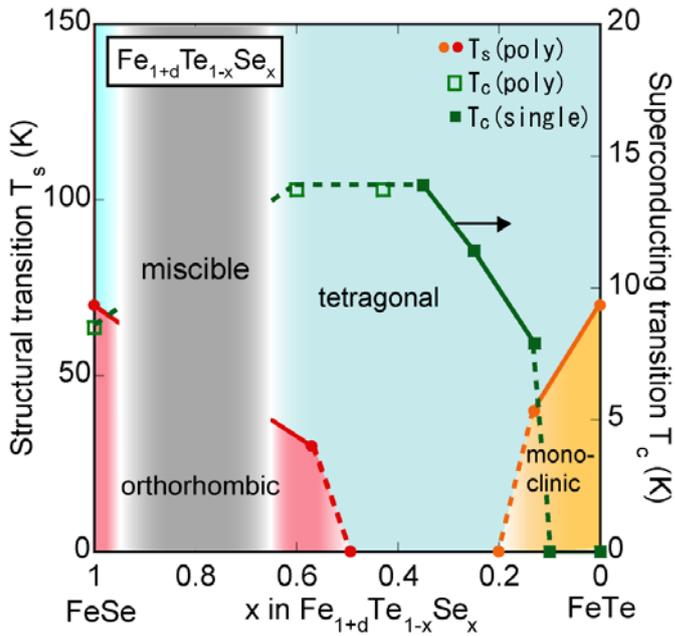

Fig. 18. Phase diagram of $Fe_{1+d}Te_{1-x}Se_x$ with low excess-Fe concentration. The tetragonal-orthorhombic structural transition is suppressed with increasing Te concentration. The highest $T_c$ appears at the tetragonal phase near $x = 0.5$. With increasing Te content, the $T_c$ decreases and the antiferromagnetic ordering accompanying the tetragonal-monoclinic distortion grows up.

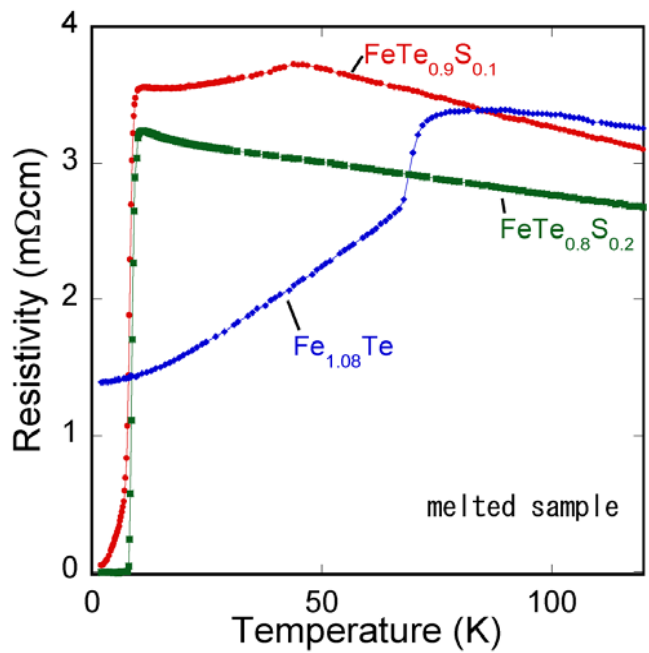

Fig. 19. Temperature dependence of resistivity for $FeTe_{1-x}S_x$.

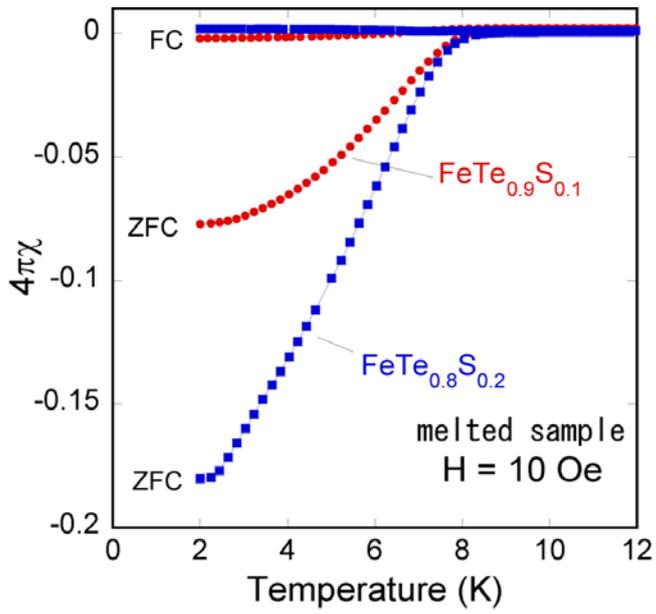

Fig. 20. Temperature dependence of magnetic susceptibility for FeTe$_{0.9}$S$_{0.1}$ and FeTe$_{0.8}$S$_{0.2}$.

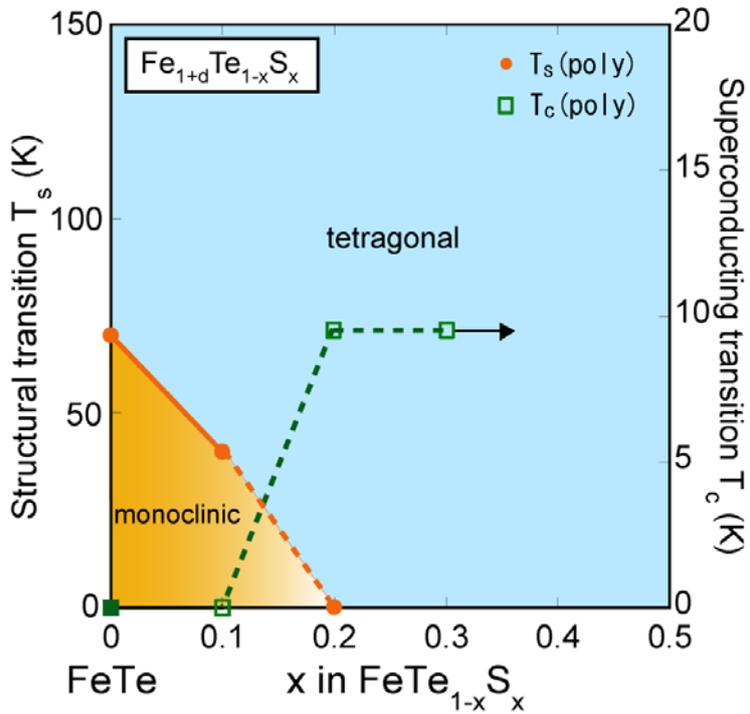

Fig. 21. Phase diagram of FeTe$_{1-x}$S$_x$ superconductor.

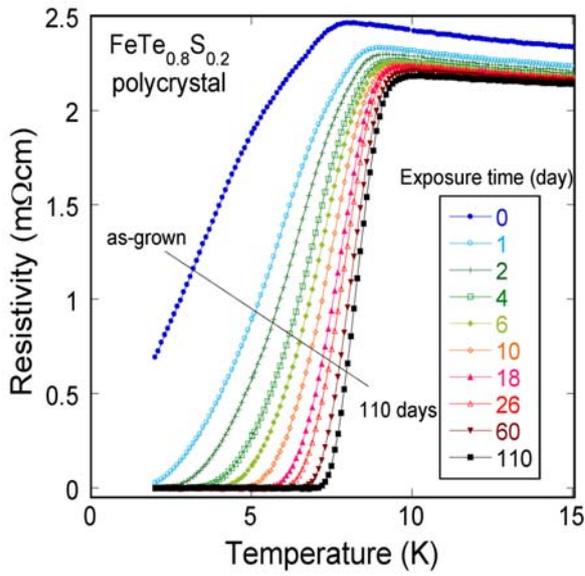

Fig. 22. Temperature dependence of resistivity for the solid-state-reacted FeTe$_{0.8}$S$_{0.2}$ sample with several air-exposure days.

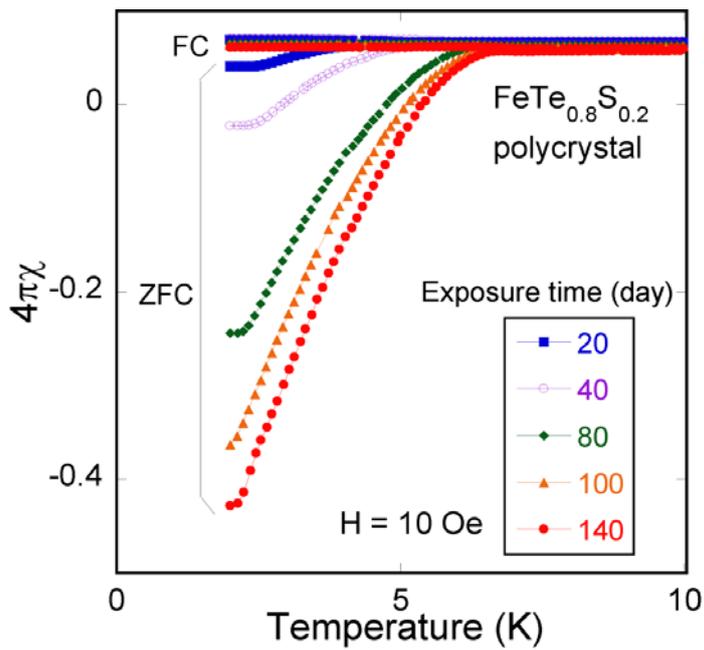

Fig. 23. Temperature dependence of magnetic susceptibility for the solid-state-reacted FeTe$_{0.8}$S$_{0.2}$ sample with several air-exposure days.

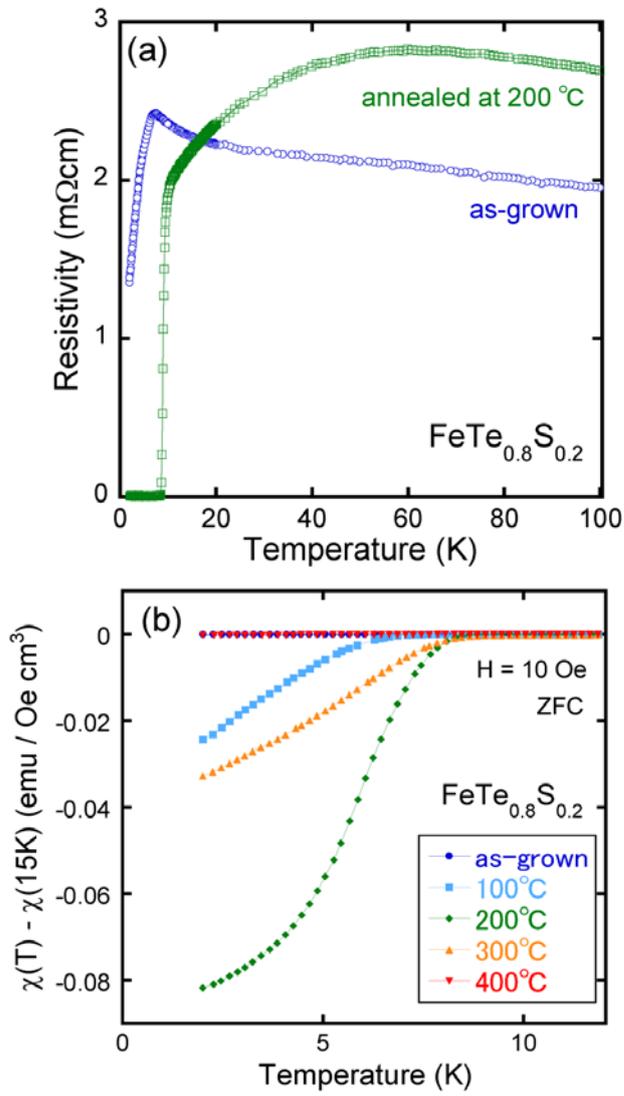

Fig. 24. (a)Temperature dependence of resistivity for as-grown FeTe$_{0.8}$S$_{0.2}$ and oxygen-annealed FeTe$_{0.8}$S$_{0.2}$.

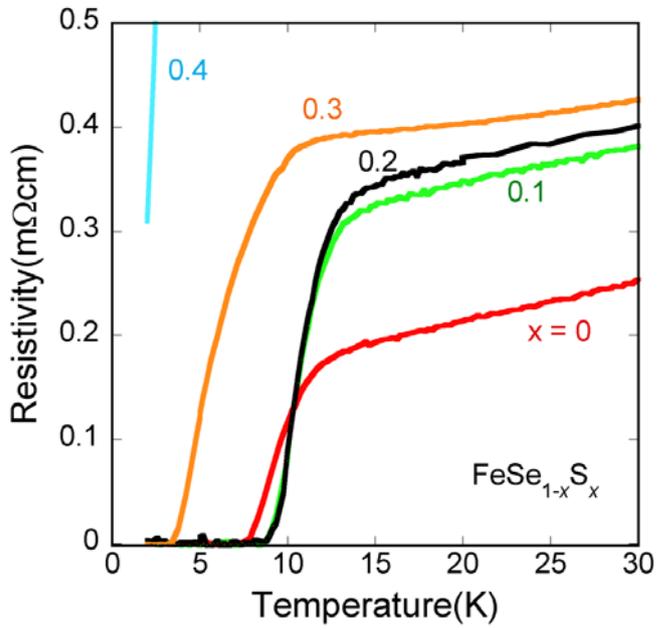

Fig. 25. Temperature dependence of resistivity for FeSe$_{1-x}$S$_x$.

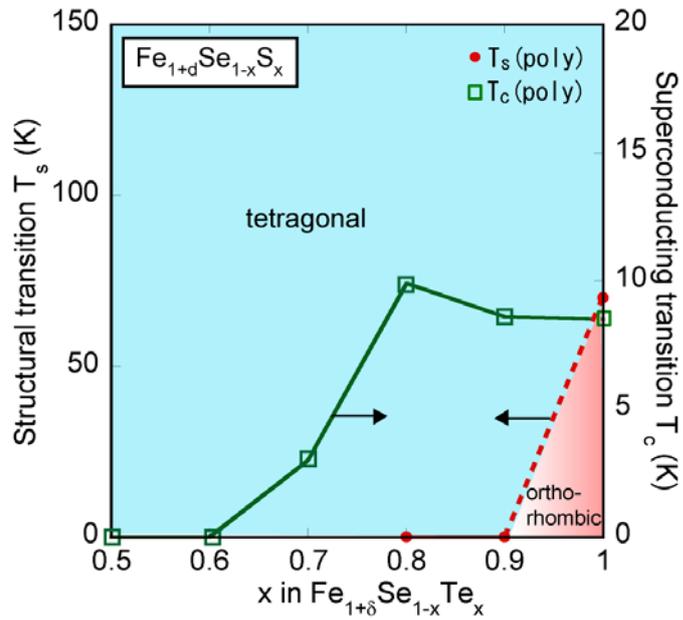

Fig. 26. Phase diagram of FeSe$_{1-x}$S$_x$.

## 3. Pressure effects of Fe chalcogenides
### 3-1. FeSe

FeSe shows the most dramatic pressure dependence of $T_c$ among the Fe chalcogenides. The $T_c^{onset}$ and $T_c^{zero}$ of FeSe at ambient pressure are 13 and 8.5 K, respectively. The $T_c^{onset}$ dramatically increased to 27 K at 1.48 GPa; the first observation of the huge pressure effect was achieved using a piston-cylinder cell, as shown in Fig. 27.[32] Interestingly, the transition became sharper around 0.5 GPa than that at ambient pressure. With applying further pressure using a diamond-anvil cell, the $T_c^{onset}$ reached 37 K as displayed in Fig. 28.[33,34] $T_c = 37$ K is the third record among the binary superconductors; the first record is 39 K of $MgB_2$, and the second record is 38 K of $Cs_3C_{60}$ under high pressure.[35,36] The precise pressure dependence of resistive transition was studied using an indenter cell as shown in Fig. 29, and the estimated $T_c^{onset}$ and $T_c^{zero}$ were plotted in Fig. 30 as a function of applied pressure.[37] Also in this measurement, the transition became sharper around 0.5 ~ 1 GPa as observed in Fig. 26. The pressure dependence of $T_c$ had an anomaly at 1-2 GPa and the $T_c$ increases up to 37 K above 2 GPa.

To investigate the origin of this huge pressure effect of FeSe, NMR under high pressure was performed.[37, 38] Figure 31 shows the temperature dependence of $1/T_1T$ for superconducting $Fe_{1.01}Se$ under high pressure up to 2.2 GPa, and for non-superconducting $Fe_{1.03}Se$. For superconducting sample, $1/T_1T$ increased with decreasing temperature above $T_c$, indicating that the superconductivity set in at $T_c$ after antiferromagnetic spin fluctuations were enhanced. Furthermore, $1/T_1T$ was enhanced with applying pressure, implying the huge pressure effect on $T_c$ in FeSe was positively linked with the enhancement of the antiferromagnetic spin fluctuations.

Crystal structural analysis under high pressure was performed for FeSe by the synchrotron x-ray diffraction.[33,34] The pressure dependence of structural parameters was summarized in Fig. 32. With increasing pressure, the lattice constants $a$, $b$, $c$, $V$ and Fe-Se distance decreased monotonously. The Se-Fe-Se angle (α angle) decreased from 104.53° (at 0.25 GPa) to 103.2° (at 9.0 GPa) with increasing pressure. The α-angle dependence of $T_c$ for FeSe deviated from what would be predicted from the Lee's plot for LaFeAsO system; the $T_c$ of LaFeAsO system depended on the α angle and the highest $T_c$ appeared near α ~ 109.47° that is the value of the regular tetrahedron.[39] Focused on the Se height from the Fe layer, the pressure dependence of Se height shows an anomaly around 1 GPa. The Se height decreased suddenly around 1 GPa, and approached to the minimum value of ~1.42 Å around 4-6 GPa. To discuss the correlation between the Se height and $T_c$, both pressure dependence of the Se height and $T_c$ were plotted in Fig. 33. Both the $T_c$ and Se height have the anomaly around 1-2 GPa, and the $T_c$ dramatically increases above this anomaly, implying the direct correlation between the $T_c$ and Se height. As displayed in the inset of Fig. 32(e), the pressure-induced orthorhombic-hexagonal transition was observed above 6 GPa. As shown in Fig. 34, showing a pressure-temperature phase diagram of FeSe, the $T_c$ began to decrease where the orthorhombic-hexagonal transition began to occur. Stabilization of the orthorhombic or tetragonal structure up to high pressure would be a key to achieve higher $T_c$ under high pressure for this system.

3-2. $FeTe_{1-x}Se_x$

We discuss the pressure effects of the mixed phase of $FeTe_{1-x}Se_x$, which has the highest $T_c$ among the Fe-chalcogenide superconductors at ambient pressure. As observed in FeSe, positive pressure effect was observed for $FeTe_{1-x}Se_x$.[24,40,41] Figure 35(a) shows the temperature dependence of resistivity for $Fe_{1.03}Te_{0.43}Se_{0.57}$ under high pressure up to 11.9 GPa. The crystal structural analysis under high pressure was also performed by the synchrotron x-ray diffraction. Figure 35(b) displays a pressure-temperature phase diagram for $Fe_{1.03}Te_{0.43}Se_{0.57}$. A pressure-induced orthorhombic-monoclinic transition was observed around 2-3 GPa, and the $T_c$ decreased above this pressure region. Also for $FeSe_{0.5}Te_{0.5}$, similar pressure dependence of $T_c$ was observed as shown in Fig. 36. Furthermore, $FeTe_{0.75}Se_{0.25}$, which is the superconductor close to the antiferromagnetically ordered phase, also showed the positive pressure effect. Figure 37 shows the temperature dependence of magnetization under high pressure up to 0.99 GPa for $FeTe_{0.75}Se_{0.25}$. With increasing pressure, both the $T_c$ and superconducting volume fraction were enhanced. These results suggest that almost of the $FeTe_{1-x}Se_x$ superconductors show the positive pressure effect on $T_c$.

To investigate the correlation between superconductivity and magnetism, the $^{125}$Te-NMR measurement was performed. As shown in Fig. 38, the $T_c$ increased accompanying an enhancement of $1/T_1T$ under high pressure, indicating that the enhancement of antiferromagnetic spin fluctuations were positively linked with the enhancement of $T_c$ also for $FeTe_{0.5}Se_{0.5}$, as observed in FeSe.

3-3. $FeTe_{1-x}S_x$ and $FeSe_{1-x}S_x$

Contrary to the cases of both FeSe and $FeTe_{1-x}Se_x$, $FeTe_{1-x}S_x$ showed a negative pressure effect on $T_c$.[42] Figure 39(a) shows the temperature dependence of resistivity for the $FeTe_{0.8}S_{0.2}$ sample whose superconductivity was induced by the air exposure as described in chapter 2. Both the $T_c^{onset}$ and $T_c^{zero}$ estimated from the resistivity measurement decreased monotonously with increasing pressure as plotted in Fig. 39(b). As described in Fig. 18 and 21, the antiferromagnetic-monoclinic phase was observed near the end member FeTe. In this respect, the negative pressure effect observed in FeTe$_{0.8}$S$_{0.2}$ might be affected by magnetism under high pressure. In fact, for FeTe$_{1-x}$Se$_{x}$, the positive pressure effect tends to be suppressed with increasing Te concentration.

Figure 40 shows the temperature dependence of magnetization for FeSe$_{0.8}$S$_{0.2}$ under high pressure up to 0.76 GPa.[41] As shown in the inset of Fig. 40, the $T_c$ did not exhibit an obvious change and the pressure dependence of $T_c$ showed a dome-shaped behavior. The dome might correspond to the low-pressure phase of FeSe, which is indicated in Fig. 33. The high-pressure studies are required for FeSe$_{1-x}$S$_{x}$.

3-4. FeTe

As mentioned in chapter 2, FeTe exhibits the antiferromagnetic ordering below 70 K and does not show superconductivity. Considered to the reports on the pressure-induced superconductivity in FeAs-based superconductors, for example AFe$_2$As$_2$ (A = Ca, Eu, Sr, Ba) system,[43-46] it is expected that FeTe would show superconductivity by an application of pressure. Figure 41 shows the temperature dependence of resistivity for Fe$_{1.08}$Te under high pressure up to 2.5 GPa.[47] The temperature at which the anomalies were observed in resistivity was plotted in Fig. 42 as a function of pressure. Although the anomaly corresponding to the antiferromagnetic ordering was suppressed, the other high-pressure phase that is not superconducting phase grew up. The structural analysis under high pressure indicated that the high-pressure phase was the collapsed tetragonal phase, which was also reported in CaFe$_2$As$_2$,[48] as shown in Fig. 42.[49] In fact, superconductivity could not be induced by applying hydrostatic pressure for FeTe. However, Han et al. recently reported that the tensile-stressed FeTe thin film showed superconductivity, suggesting the importance of the uniaxiality of the pressure to realize superconducting state.[50] The detailed will be described in chapter 5.

3-5. Anion height dependence of $T_c$

Several reports indicated that the superconducting properties of the Fe-based superconductors were correlated to their crystal structure. Here we focused on the anion height from the Fe layer as a probe to investigate the $T_c$ of Fe-based superconductors. This was motivated by both a theoretical study suggesting the anion height as "a possible switch between nodeless high-$T_c$ and nodal low-$T_c$ pairings",[51] and the experimental results for FeSe shown in Fig. 33. Figure 43(a) is the anion height dependence of $T_c$ of the typical Fe-based superconductors. A schematic image of anion height from the Fe layer was described in Fig. 43(b). The data points were selected with a following policy; the valence of Fe should be close to 2+, and the $T_c$ is the highest in that system, as described detailed in Ref. 42. The anion height dependence of $T_c$ showed a symmetric curve with a peak around 1.38 Å, as indicated by the hand-fitting curve. All of the data points obeyed the unique curve for not only ambient pressure but also under high pressure.

We discuss the pressure effects of Fe-chalcogenide superconductors using this plot. Surprisingly, the data points of FeSe under high pressure obeyed the unique curve above ~2 GPa. In the respect that the data points of FeSe obeyed the unique curve above 2 GPa, an intrinsic superconductivity might be induced by the application of pressures above 2 GPa.

Focused on FeTe$_{0.43}$Se$_{0.57}$, the data point at ambient pressure is located near the unique curve. If the data points of FeTe$_{0.43}$Se$_{0.57}$ under high pressure also obey the curve, the anion (Se/Te) height should decrease from 1.620 to ~1.45 Å when the $T_c$ reached 23 K. In fact, however, the anion height at the optimal pressure was 1.598 Å that is much smaller than that expected from the unique curve, indicating that the plot was not applicable for the pressure effect of FeTe$_{1-x}$Se$_x$. One of the obvious differences between FeSe and FeTe$_{1-x}$Se$_x$ is whether the disorder exists at the anion site or not. In fact, the high-resolution x-ray single crystal diffraction for the FeTe$_{0.56}$Se$_{0.44}$ indicated the existence of significantly different anion heights of Te and Se with a differential $\Delta h_{\text{Te-Se}}$ = 0.24 Å.[52] In the concept that the $T_c$ of the Fe-based superconductor strongly depends on the anion height, the disorder at the anion site should strongly affect the superconducting properties of the Fe-based superconductor. In this respect, understanding of the mechanism of the pressure effect of the Fe-based superconductors that contains the disorder at the anion site will require more detailed microscopic investigations.

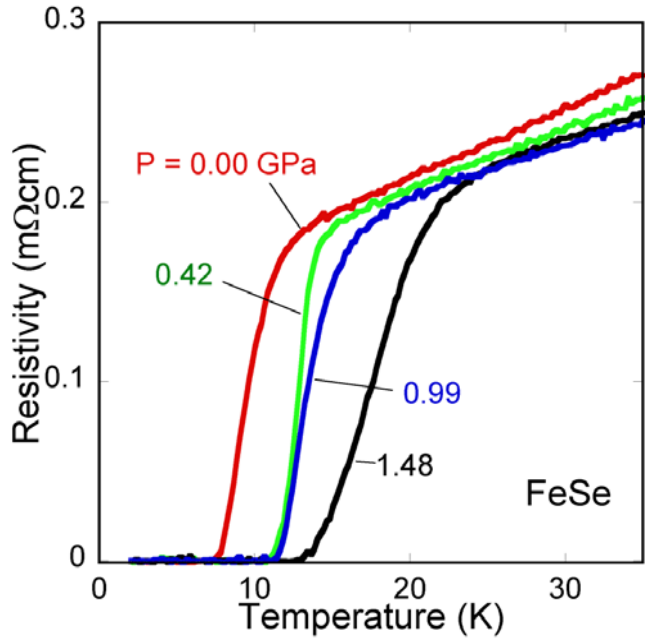

Fig. 27. Temperature dependence of resistivity for FeSe under high pressure up to 1.48 GPa.

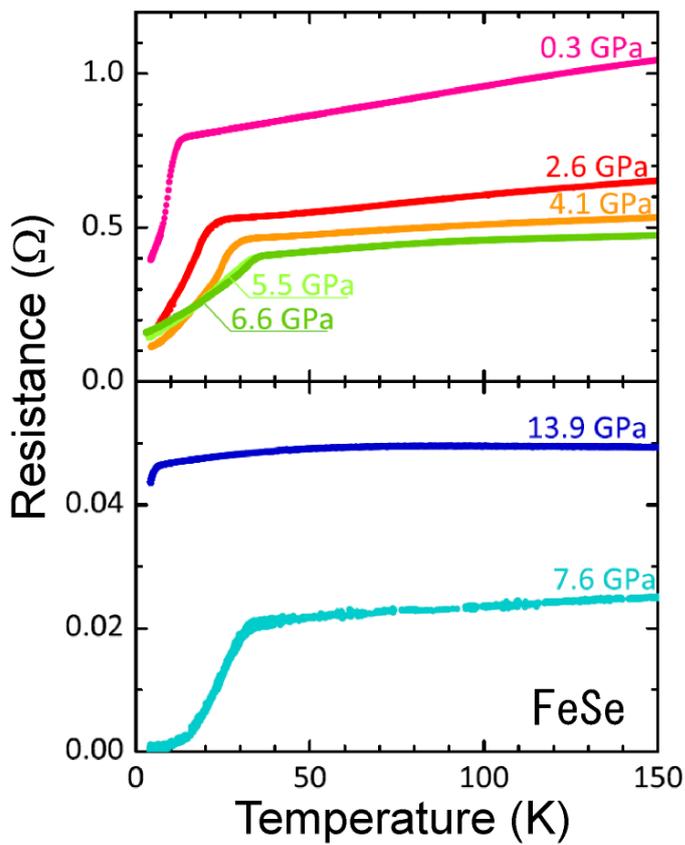

Fig. 28. Temperature dependence of resistivity measured using a diamond-anvil cell for FeSe under high pressure up to 13.9 GPa.
[Figures reprinted from S. Margadonna et al., Phys. Rev. B 80, 064506 (2009).
Copyright 2009 by The American Physical Society.]

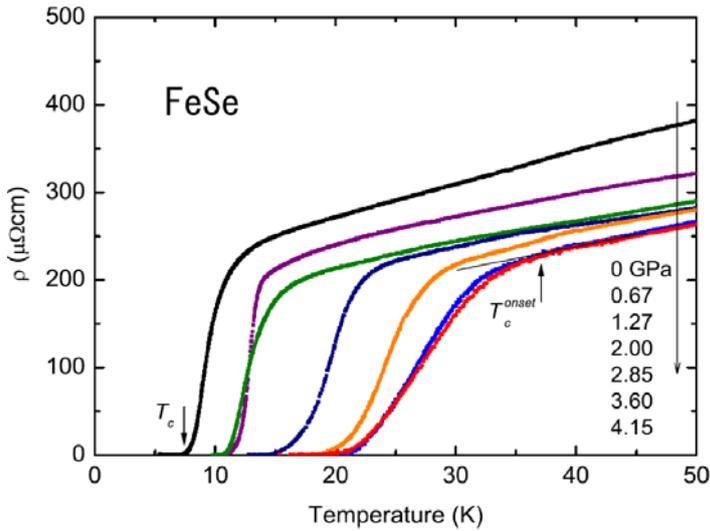

Fig. 29. Temperature dependence of resistivity measured using an indenter cell for FeSe under high pressure up to 4.15 GPa.
[Figures reprinted from S. Masaki et al., J. Phys. Soc. Jpn. 78, 063704 (2009).
Copyright 2009 by Physical Society of Japan.]

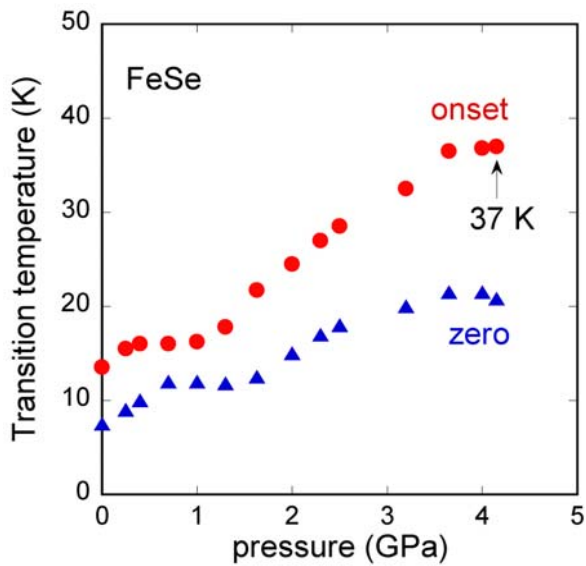

Fig. 30. Pressure dependence of $T_c^{zero}$ and $T_c^{onset}$ for FeSe.

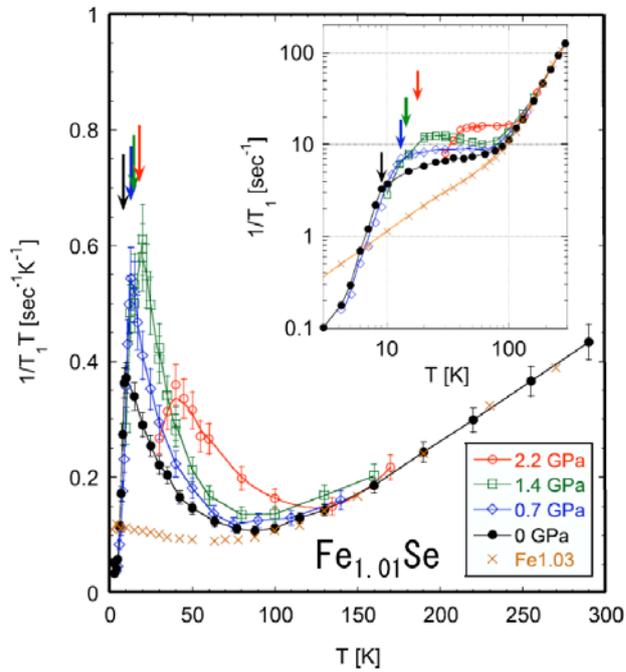

Fig. 31. $1/T_1T$ for superconducting $Fe_{1.01}Se$ under various pressures, and for non-superconducting $Fe_{1.03}Se$ in $P = 0$. $1/T_1T$ reflects the spin fluctuation susceptibility averaged over various wave-vector modes $q$. Inset: log-log plot of $1/T_1$. Vertical arrows mark (from left to right) $T_c$ for 0, 0.7, 1.4 and 2.2 GPa.
[Figures reprinted from T. Imai et al., Phys. Rev. Lett 102, 177005 (2009). Copyright 2009 by The American Physical Society.]

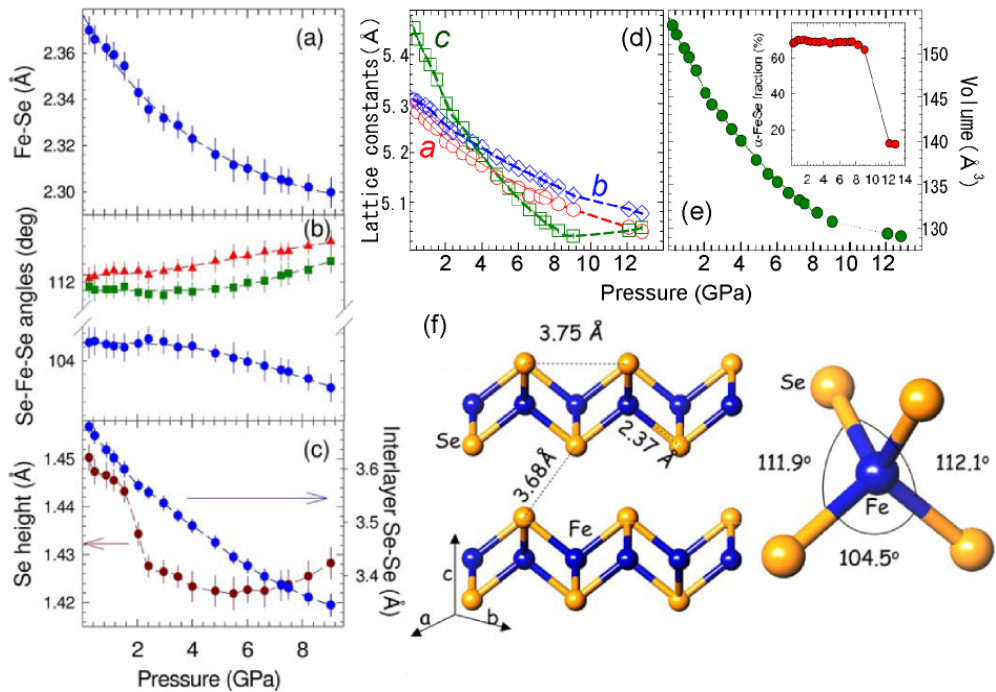

Fig. 32. Crystal structural parameters of FeSe under high pressure: (a) Fe-Se distance, (b) Se-Fe-Se angle, (c) Se height from Fe layer, (d) lattice constants $a$, $b$ and $c$, (e) Volume, (f) crystal structure of FeSe. The inset in (e) shows the pressure dependence of α-FeSe (orthorhombic FeSe) fraction.
[Figures reprinted from S. Margadonna et al., Phys. Rev. B 80, 064506 (2009). Copyright 2009 by The American Physical Society.]

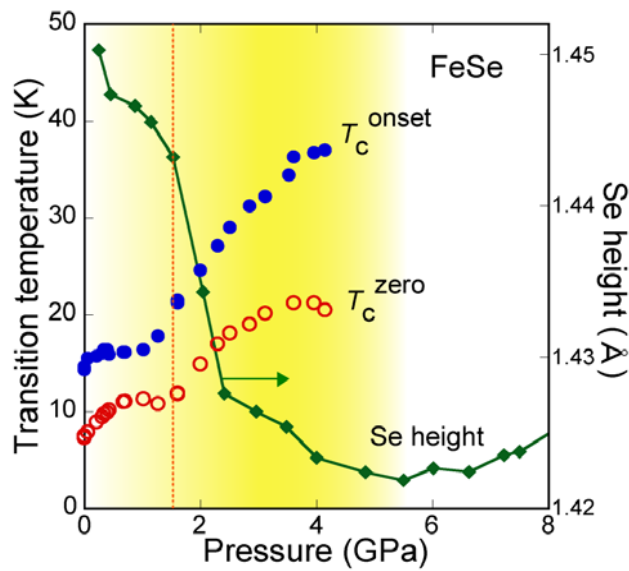

Fig. 33. Pressure dependence of $T_c$ and Se height from the Fe layer. Both $T_c$ and Se height have the anomaly around 1-2 GPa.

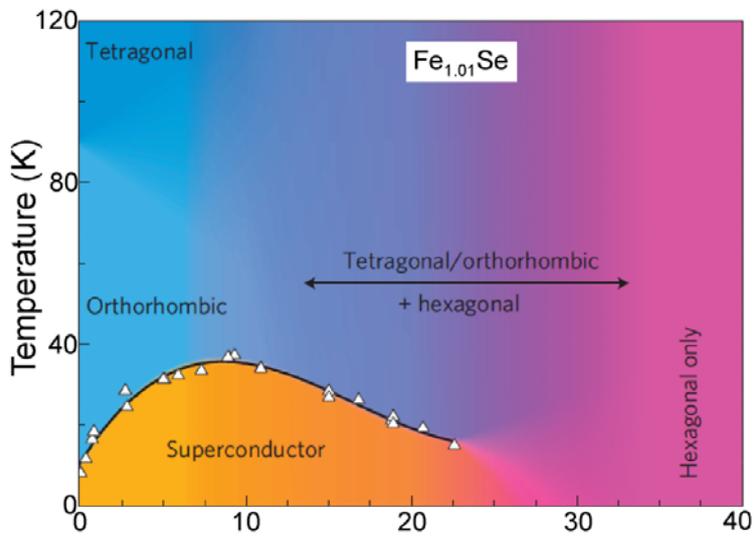

Fig. 34. Pressure-temperature phase diagram of $Fe_{1.01}Se$.
[Figure reprinted from S. Medvedev et al., Nat. Mater. 8, 630 (2009). Copyright 2009 by Macmillan Publishers Limited.]

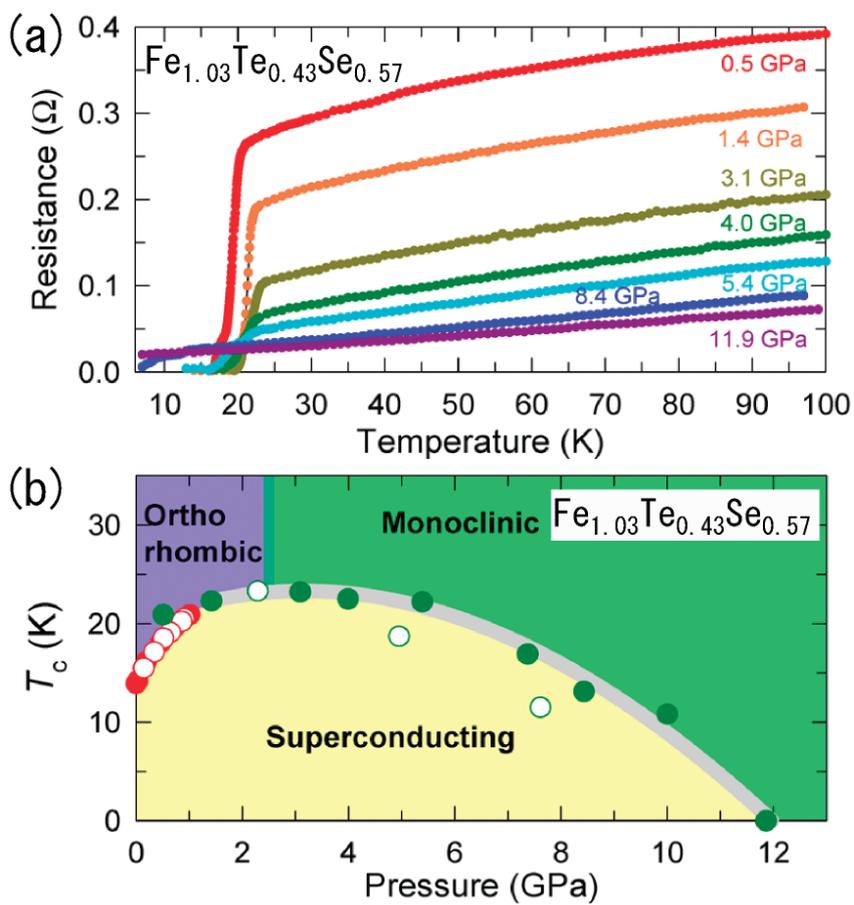

Fig. 35. (a) Temperature dependence of resistivity for $Fe_{1.03}Te_{0.43}Se_{0.57}$ under high pressure up to 11.9 GPa. (b) Pressure-temperature phase diagram of $Fe_{1.03}Te_{0.43}Se_{0.57}$.
[Figures reprinted from N. C. Gresty et al., J. Am. Chem. Soc. 131, 16944 (2009). Copyright 2009 by American Chemical Society.]

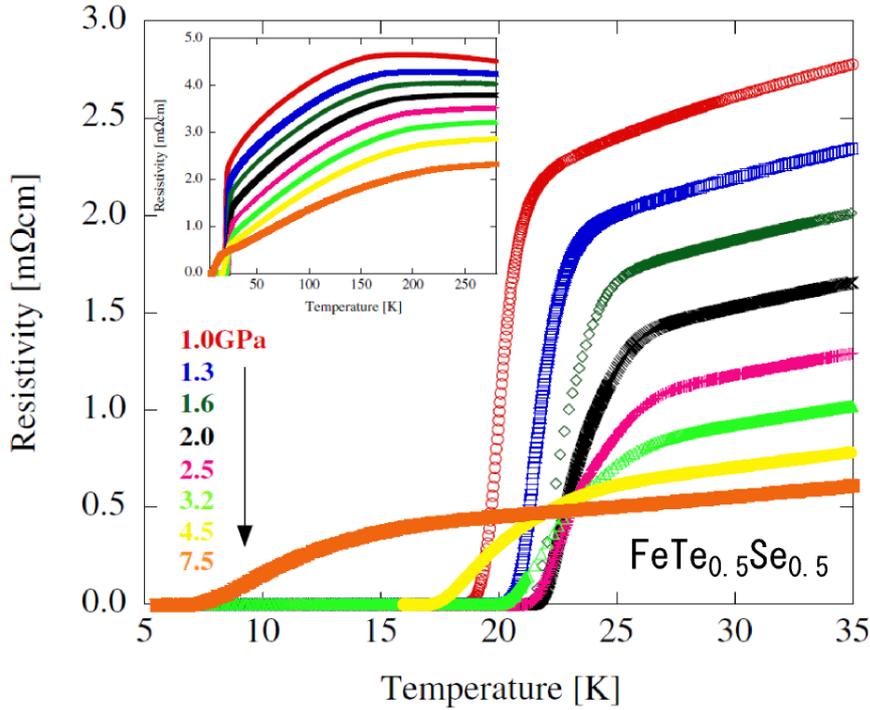

Fig. 36 Temperature dependence of resistivity for $FeTe_{0.5}Se_{0.5}$ under high pressure up to 7.5 GPa.
[Figure reprinted from K. Horigane et al., J. Phys. Soc. Jpn. 78, 063705 (2009). Copyright 2009 by The Physical Society of Japan.]

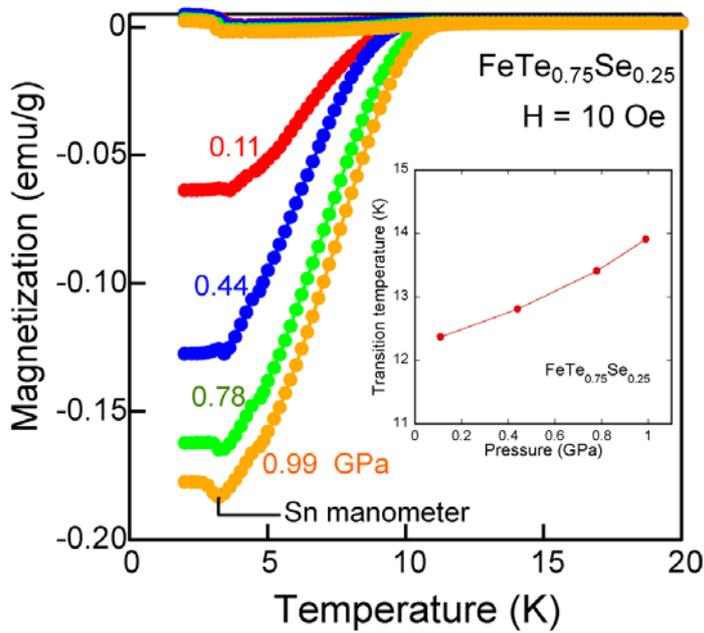

Fig. 37. Temperature dependence of magnetization for FeTe$_{0.75}$Se$_{0.25}$ under high pressure up to 0.99 GPa. The inset shows the pressure dependence of $T_c$ estimated from the magnetization measurement.

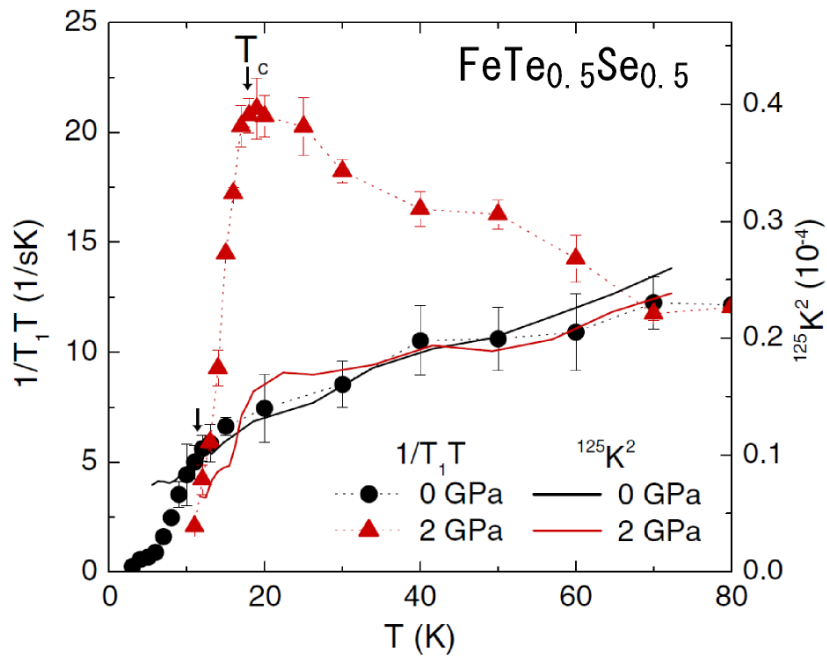

Fig. 38. Temperature dependence of $1/T_1T$ and $^{125}K^2$ in FeTe$_{0.5}$Se$_{0.5}$.
[Figure reprinted from Y. Shimizu et al., J. Phys. Soc. Jpn. 78, 123709 (2009). Copyright 2009 by The Physical Society of Japan.]

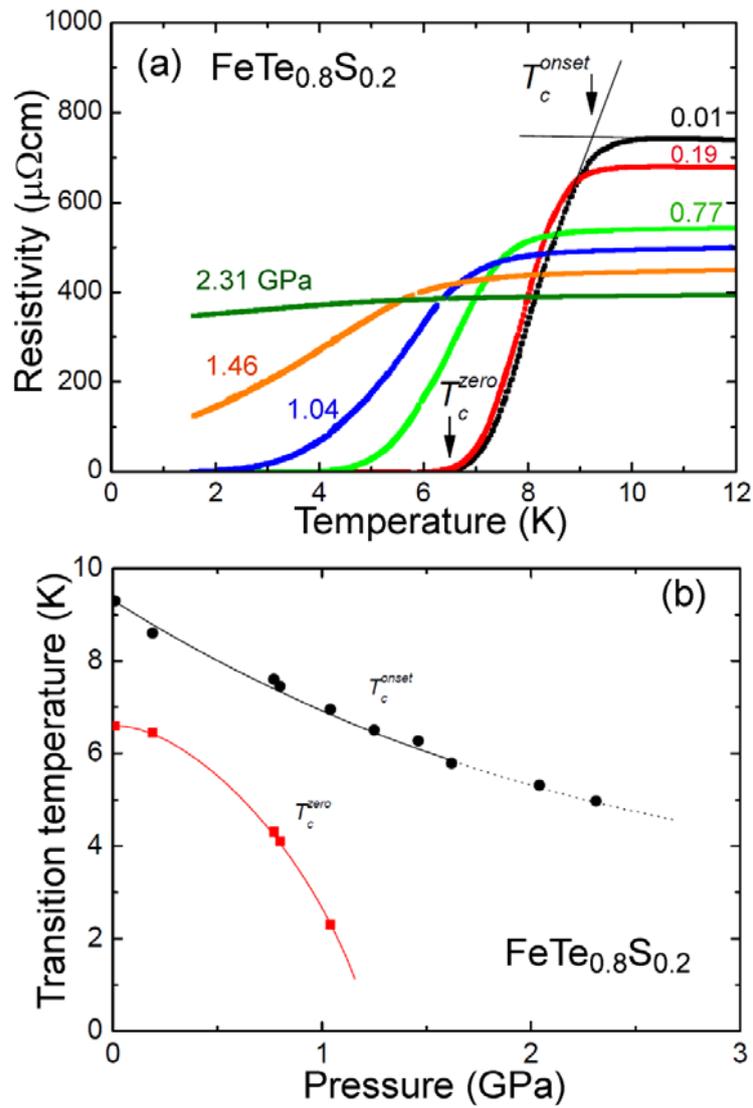

Fig. 39 (a) Temperature dependence of resistivity for FeTe$_{0.8}$S$_{0.2}$ under high pressure up to 2.31 GPa. (b) Pressure dependence of $T_c^{onset}$ and $T_c^{zero}$ for FeTe$_{0.8}$S$_{0.2}$.

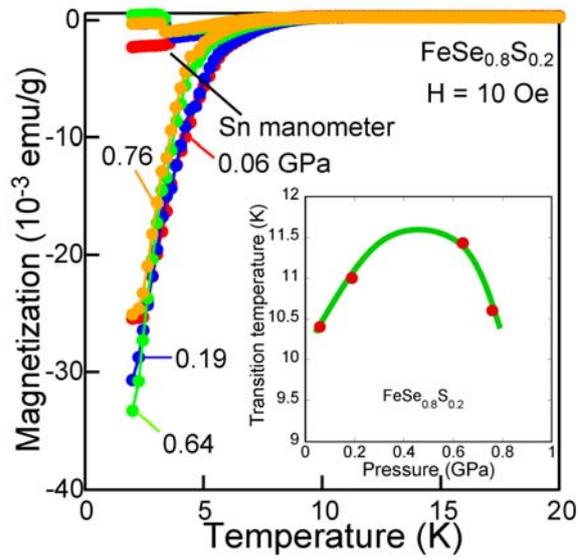

Fig. 40. Temperature dependence of magnetization for FeSe$_{0.8}$S$_{0.2}$ under high pressure up to 0.76 GPa. The inset shows the pressure dependence of $T_c$ estimated from the magnetization measurement for FeSe$_{0.8}$S$_{0.2}$.

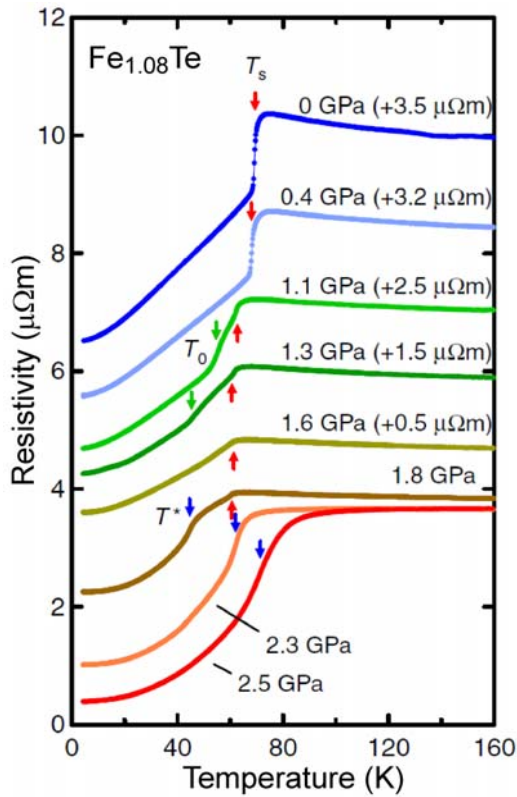

Fig. 41. Temperature dependence of resistivity for Fe$_{1.08}$Te under high pressure up to 2.5 GPa.
[Figure reprinted H. Okada et al., J. Phys. Soc. Jpn. 78, 083709 (2009). Copyright 2009

by The Physical Society of Japan.]

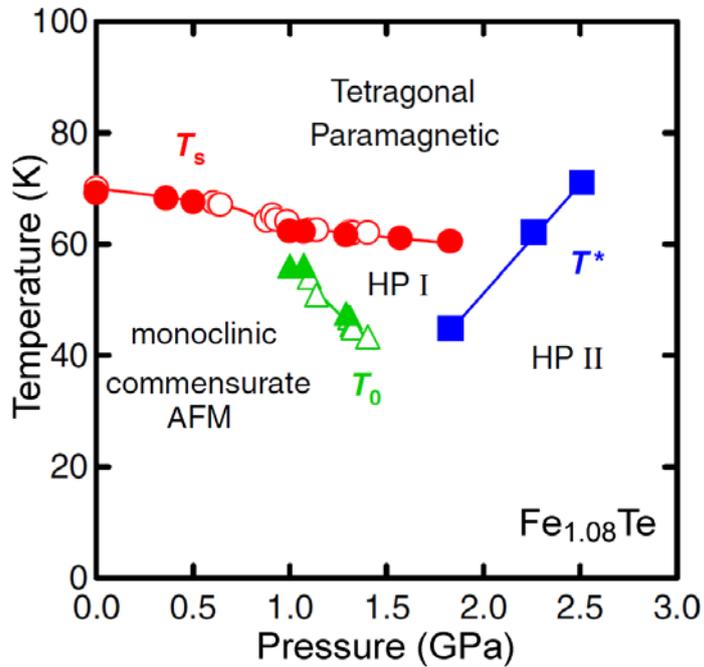

Fig. 41. Pressure-temperature phase diagram of $Fe_{1.08}Te$.
[Figure reprinted H. Okada et al., J. Phys. Soc. Jpn. 78, 083709 (2009). Copyright 2009 by The Physical Society of Japan.]

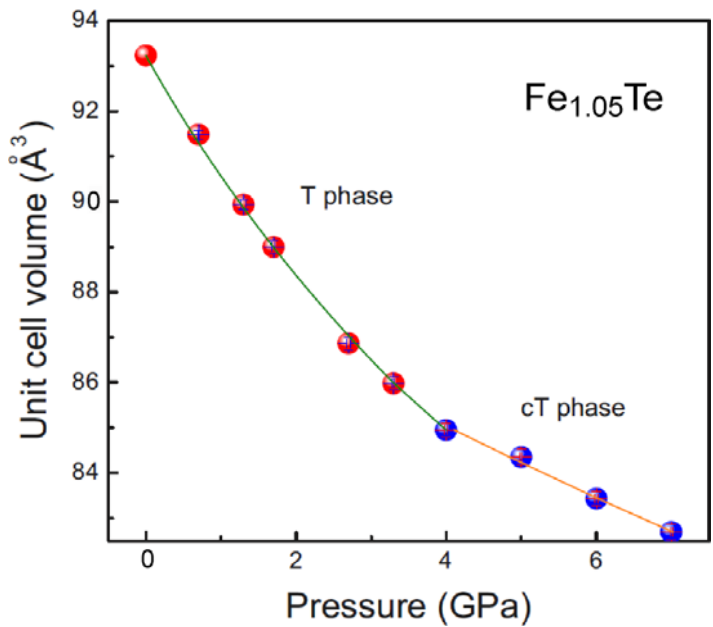

Fig. 42. Pressure dependence of unit cell volume at 300 K for $Fe_{1.05}Te$. The high-pressure phase is the collapsed tetragonal phase.

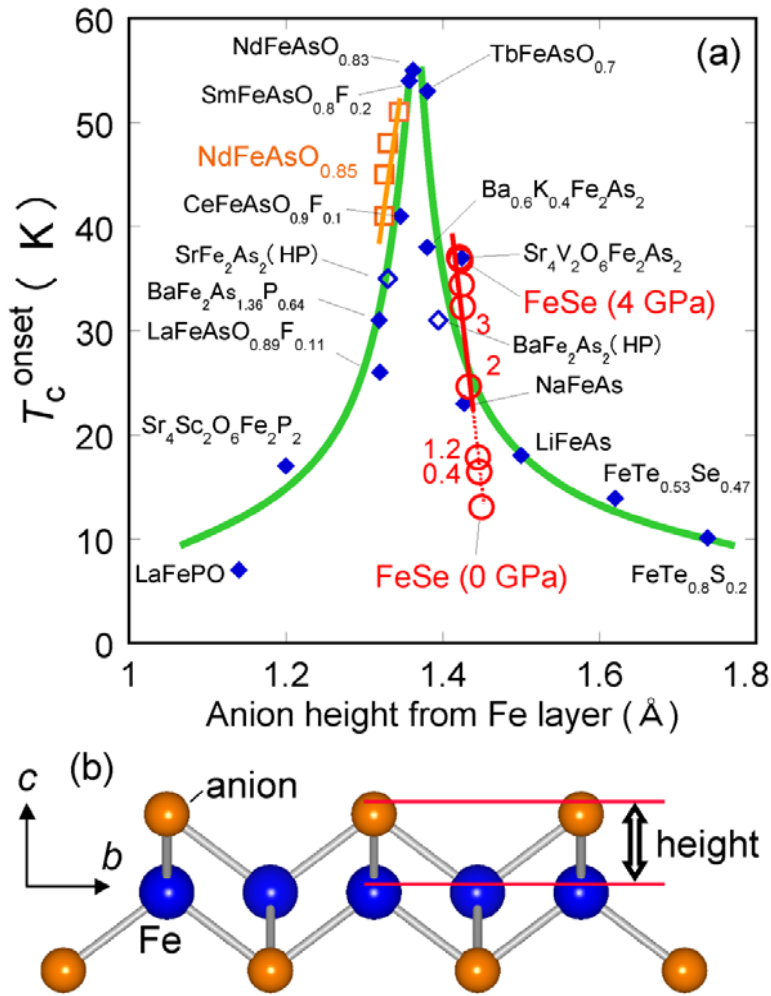

Fig. 43. (a) Anion height dependence of $T_c$ of the typical Fe-based superconductors. Filled and open marks indicate the data points at ambient pressure and under pressure, respectively. (b) Schematic image of the anion height from the Fe layer.

4. Superconducting gap

Investigations of the superconducting gap are essential for both theoretical and experimental studies to elucidate the mechanism of superconductivity. The direct observation of the superconducting gap is one of the most effective ways. By the photoemission spectroscopy, the electronic structure was investigated.[53] Figure 44 shows the angle-resolved photoemission spectroscopy (ARPES) intensity at $E_F$ as a function of the two-dimensional wave vector for the $Fe_{1.03}Te_{0.7}Se_{0.3}$ single crystal. The observed Fermi surfaces were similar to that of $BaFe_2As_2$ system. The absence of the Fermi surface at the X point observed in antiferromagnetically ordered FeTe is consistent with the disappearance of long-range antiferromagnetic ordering for superconducting $Fe_{1.03}Te_{0.7}Se_{0.3}$. Also the superconducting gap was observed for $Fe_{1.03}Te_{0.7}Se_{0.3}$ by the ultrahigh-resolution ARPES measurements in the close vicinity of $E_F$, and the observed and symmetrized spectra were described in Fig. 45. The estimated superconducting gap size ($\Delta \sim 4$ meV) corresponded to a $2\Delta/k_BT_c$ value of ~7, indicating the strong-coupling nature of superconductivity in $Fe_{1.03}Te_{0.7}Se_{0.3}$.

The large $2\Delta/k_BT_c$ value was also indicated by both the high-field transport measurements for $FeTe_{0.75}Se_{0.25}$ and NMR for $FeTe_{0.8}Se_{0.2}$. Figure 46 shows the temperature dependence of $1/T_1$ for FeSe (Se-NMR) and $Fe_{1.086}Te_{0.8}Se_{0.2}$ (Te-NMR).[54,55] For $Fe_{1.086}Te_{0.8}Se_{0.2}$, $1/T_1$ showed $T^6$ behavior below $T_c$, indicating that the $2\Delta/k_BT_c$ value was larger than that predicted by the BCS theory. Figure 47 shows the field-temperature phase diagram for FeSe and $FeTe_{0.75}Se_{0.25}$. While the upper critical field $\mu_0H_{c2}(T)$ of FeSe obeyed the WHH theory, $\mu_0H_{c2}(T)$ of $FeTe_{0.75}Se_{0.25}$ exhibited a saturation at low temperature. The data points were fitted using a large Maki parameter of $\alpha_1 = 2.5$. This value also suggested the large $2\Delta/k_BT_c$ value of ~7.5.[56-58] Recently, a similar result was obtained also for the $FeTe_{1-x}S_x$ system.[59]

The superconducting gap was observed also by the STS measurement using the $Fe_{1.05}Te_{0.85}Se_{0.15}$ single crystal.[60] Figure 48 shows the spatially averaged spectrum at 4.2 K normalized to the background conductance for $Fe_{1.05}Te_{0.85}Se_{0.15}$ with a fit of the calculated DOS for the *s*-wave superconductor. As shown in Fig. 49, the magnitude of the gap does not show significant spatial variation, in spite of the existence of the excess Fe. The $\Delta$ value was estimated to be 3meV. By using the $T_c = 8.0$ K estimated from the phase diagram of $FeTe_{1-x}Se_x$ (Fig. 18), the $2\Delta/k_BT_c$ value was calculated to be 6.65, also suggesting the $Fe_{1.05}Te_{0.85}Se_{0.15}$ would be the strong-coupling superconductor.

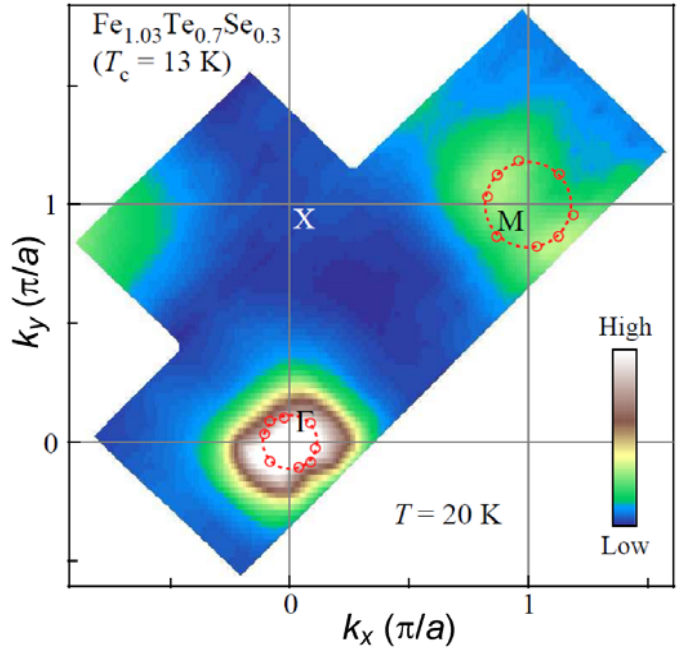

Fig. 44. ARPES intensity plot at EF of $Fe_{1.03}Te_{0.7}Se_{0.3}$ as a function of the two-dimensional wave vector measured at 20 K with 44 eV photons. The intensity at $E_F$ is obtained by integrating the spectra within ±10 meV with respect to $E_F$. Solid and dashed red circles show experimentally determined $k_F$ points and schematic FSs, respectively. There are sizable experimental uncertainties on the experimentally determined $k_F$ points, mainly due to weak intensity around the M point.
[Figure reprinted from K. Nakayama et al., arXiv:0907.0763.]

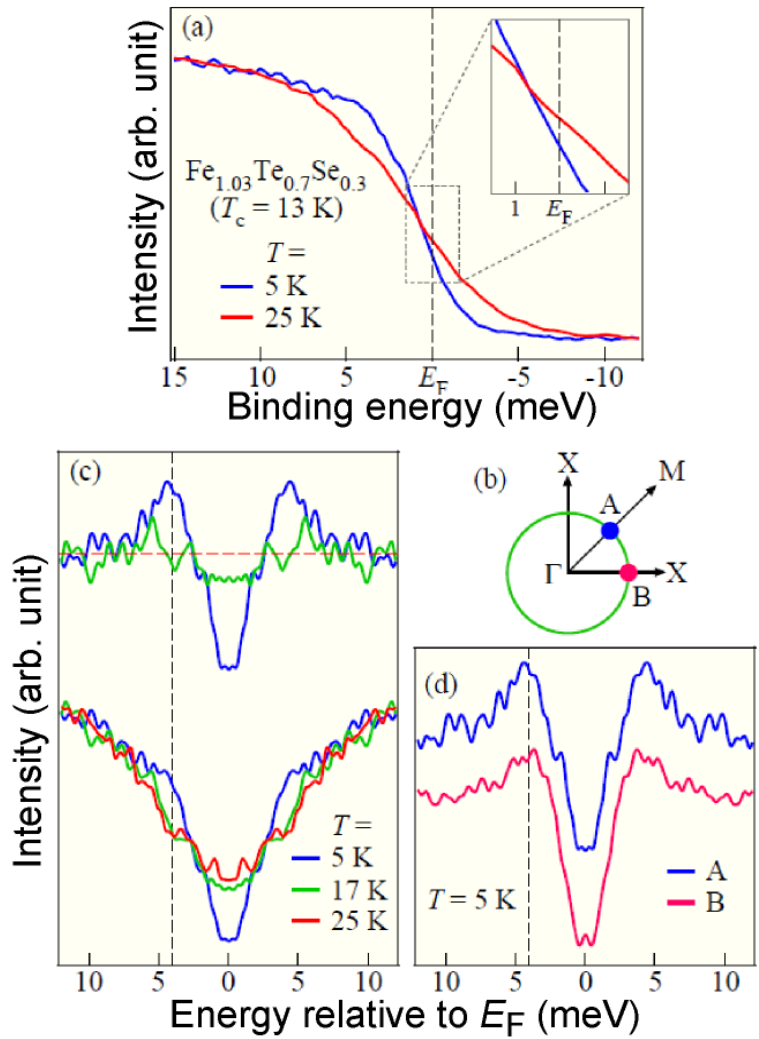

Fig. 45. (a) Ultrahigh-resolution ARPES spectra of $Fe_{1.03}Te_{0.7}Se_{0.3}$ near $E_F$ at 5 and 25 K (blue and red curve, respectively) with the He I$\alpha$ resonance line, measured at point A of the outer hole pocket displayed in (b). The inset shows the expansion in the vicinity of $E_F$. (b) Schematic hole-like Fermi surface at the $\Gamma$ point with the location of the $k_F$ points A and B. (c) Temperature dependence of symmetrized ARPES spectra (bottom) at point A, and the same but divided by the spectrum at 25 K (top). (d) Comparison of the symmetrized spectra at $k_F$ points along $\Gamma$-M (blue curve) and $\Gamma$-X (purple curve) high-symmetry lines measured at 5 K divided by the 25 K spectrum. Dashed lines at 4 meV in (c) and (d) represent the energy scale of the superconducting gap $\Delta$.
[Figure reprinted from K. Nakayama et al., arXiv:0907.0763.]

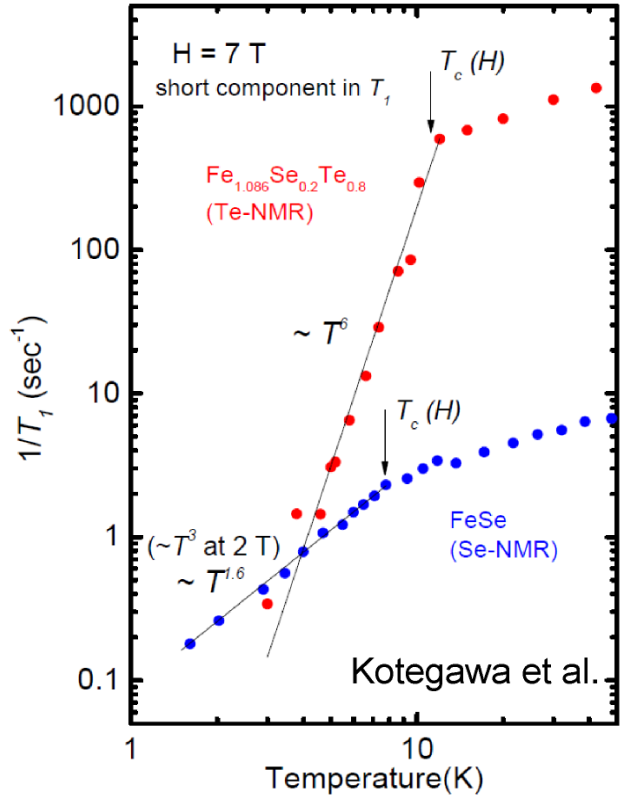

Fig. 46. Temperature dependence of $1/T_1$ for FeSe (Se-NMR) and $Fe_{1.086}Te_{0.8}Se_{0.2}$ (Te-NMR). This results will be published by H. Kotegawa et al.

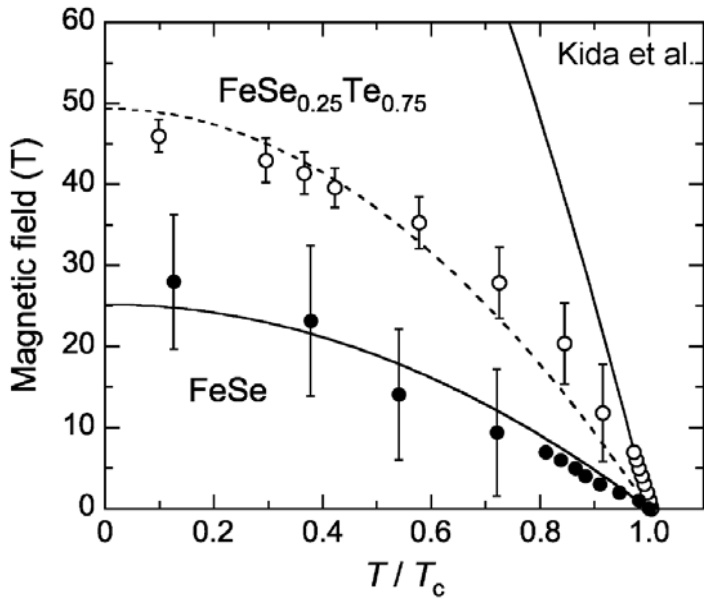

Fig. 47. Field-temperature phase diagram for FeSe and $FeTe_{0.75}Se_{0.25}$. The solid lines indicate the magnetic fields calculated using the WHH theory. The dash line shows the fitted line with a large Maki parameter of $\alpha_1 = 2.5$.

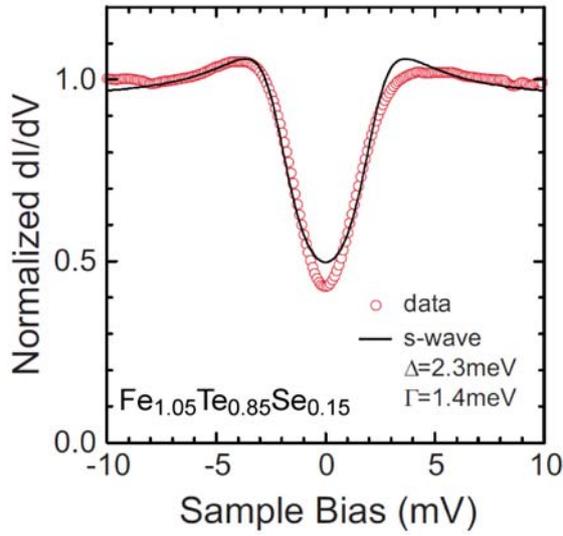

Fig.48. Spatially averaged spectrum at 4.2 K normalized to the background conductance for $Fe_{1.05}Te_{0.85}Se_{0.15}$. The curve indicates a fit of the calculated DOS for an *s*-wave superconductor to the data.
[Figure reprinted from T. Kato et al., Phys Rev. B 80, 180507 (2009). Copyright 2009 by The American Physical Society.]

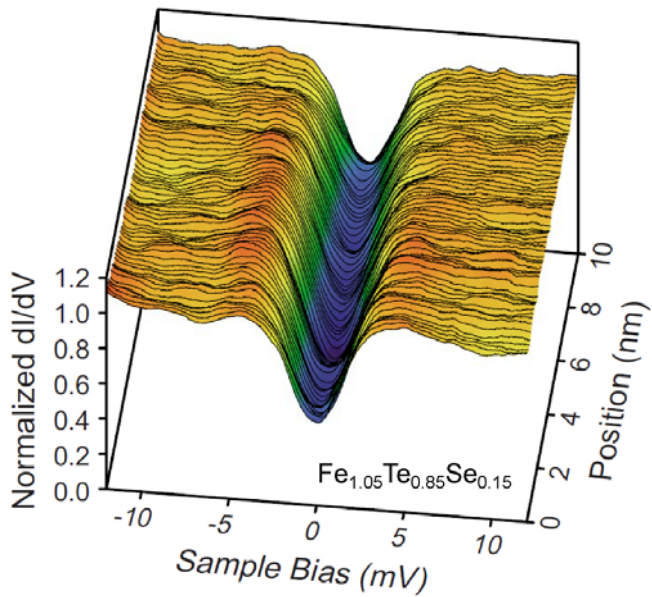

Fig. 49. Normalized spectra at 4.2 K taken along a 10-nm-long linecut across the sample surface. All of the raw spectra were normalized to the background conductance.
[Figure reprinted from T. Kato et al., Phys Rev. B 80, 180507 (2009). Copyright 2009 by The American Physical Society.]

5. Fabrications of thin film and superconducting wire using Fe chalcogenides
5-1. Thin film

Fe-chalcogenide superconductor is advantageous for fabrication of both thin films and superconducting wires, because it is basically binary alloy. To date, the superconducting thin films have been fabricated for FeSe, FeTe, FeTe$_{1-x}$Se$_x$ and FeTe$_{1-x}$S$_x$.[50,61-66] Figure 50 shows the temperature dependence of resistivity for FeSe film prepared with two substrate temperatures of a low temperature (LT) = 320 ºC and a high temperature (HT) = 500 ºC with several different thickness. For LT-FeSe, the onset of the superconducting transition appeared with the thickness of 140nm. The $T_c^{onset}$ increased with increasing thickness. The appearance of superconductivity seems to require enough thickness above 140 nm. On the other hand, HT-FeSe film did not exhibit such thickness dependence of $T_c$. The lattice distortion observed in the bulk sample was not observed in the LT-film that did not show superconductivity, while HT-film, which showed the sharp superconducting transition, showed the lattice distortion. This difference suggests that the appearance of superconductivity requires lattice distortion observed at 70-90 K in the bulk sample.[19,20] Not only the thin film but also the nano sheets of FeSe and FeTe$_{1-x}$Se$_x$ with a thickness of ~2 nm were synthesized. However, the nsno sheets did not show superconductivity. To clarify the relationship between lattice distortion and superconductivity, more detailed investigations on the thin films and nano sheets should be addressed for the FeSe system.

Tensile-stress effects on $T_c$ was reported for FeTe$_{1-x}$Se$_x$.[64,65] The $T_c$ of FeTe$_{1-x}$Se$_x$ film increased with decreasing lattice constant $a$, and reached 21 K as shown in Fig. 51. In fact, the tensile stress can raise $T_c$ in Fe-chalcogenide superconductors, as physical pressure achieved. Surprisingly, the FeTe film which was strain-stressed showed superconductivity without Se or S substitutions.[50] Figure 52 shows the temperature dependence of resistivity for bulk sample and FeTe thin film deposited on MgO substrate. For the strain-stressed thin film, the anomaly corresponding to the magnetic transition was suppressed and superconductivity appeared at 13 K, while the application of the hydrostatic physical pressure could not induce superconductivity in FeTe.[47]

As mentioned in chapter 3, Fe-chalcogenide superconductors are much sensitive to applying pressure. Based on the fact, we expect $T_c$ = 37 K for optimally stressed FeSe film, because the bulk sample shows superconducting transition at 37 K under optimal pressure. The tuning of lattice constants of the thin films will realize high $T_c$ superconductivity in Fe chalcogenides at ambient pressure.

5-2. Superconducting wire

The fabrication of the superconducting wire using Fe-based superconductor is one of the key challenges to investigate the possibility of the application of the Fe-based superconductors. Because of the low anisotropy of FeTe$_{1-x}$Se$_x$,[58] we tried the fabrication of the FeTe$_{1-x}$Se$_x$ superconducting wire by the powder-in-tube method using an Fe sheath and only the TeSe powder.[68] Figure 53 displays the scanning electron microscope (SEM) image of the cross section of the FeTe$_{1-x}$Se$_x$ wire. Good connections between the Fe sheath and FeTe$_{1-x}$Se$_x$ superconducting phase were obtained at the edge of the cross section. Furthermore the zero-resistivity current was observed in the current-voltage measurement. The estimated critical current density was plotted in Fig. 54. While the estimated value was small as the superconducting wire available in real application, this was the first report of the estimation of the critical current density for the Fe-based superconducting wire.

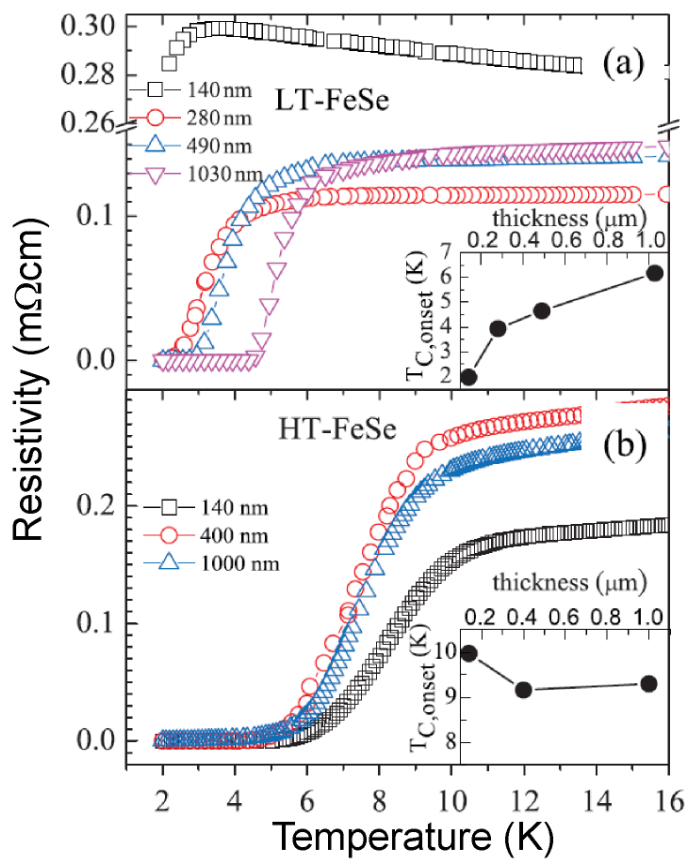

Fig. 50. (a) Temperature dependence of resistivity for LT-FeSe prepared with a low substrate temperature of 320 ºC. (b) Temperature dependence of resistivity for HT-FeSe prepared with a high substrate temperature of 500 ºC.

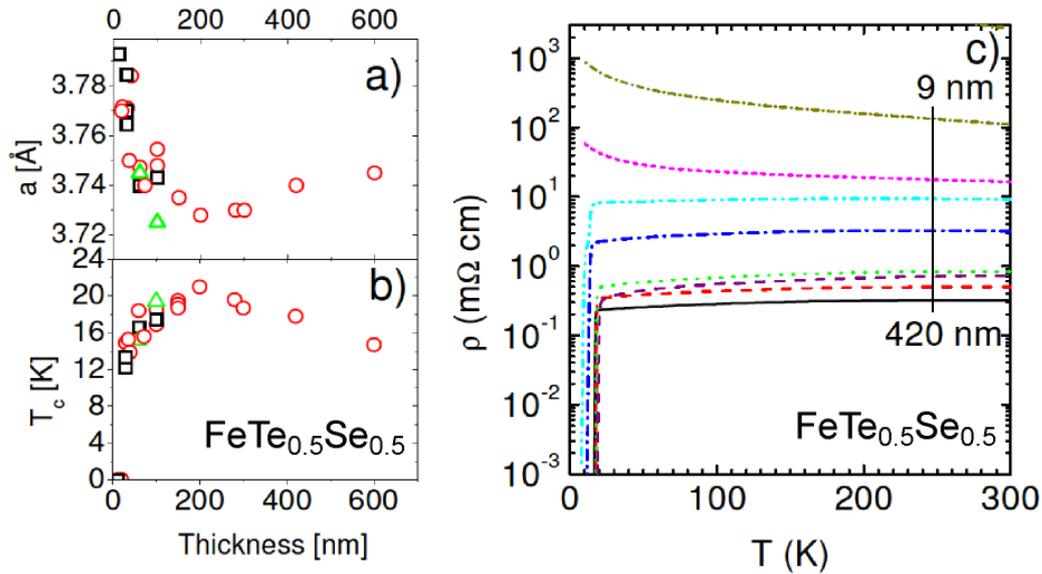

Fig. 51. (a) Thickness dependence of lattice constant $a$ for FeTe$_{1-x}$Se$_x$ films with several different thickness (9, 18, 36, 72, 150, 200, 280 and 420 nm) (b) Thickness dependence of $T_c$. The circles, squares and triangles in (a) and (b) indicate the sample deposited on LAO, STO and ZrO:Y, respectively. (c) Temperature dependence of resistivity for FeTe$_{1-x}$Se$_x$ films with several different thicknesses.

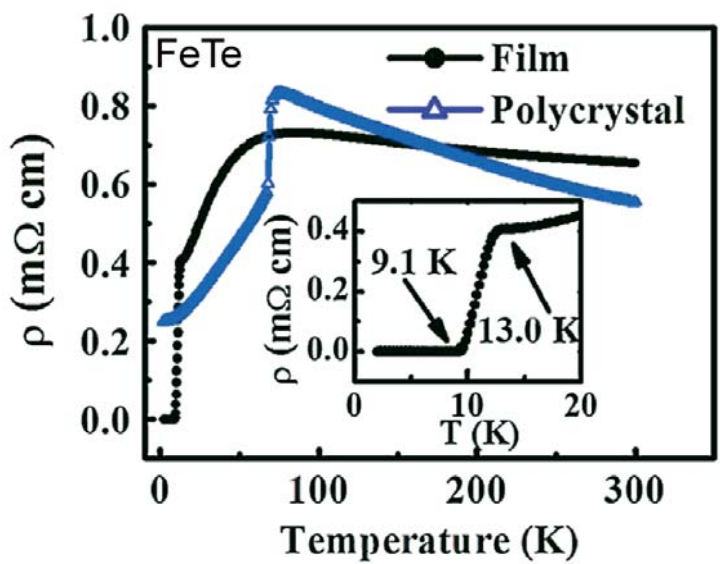

Fig. 52. Temperature dependence of resistivity for the polycrystal and thin film on MgO

substrate of FeTe.

[Figure reprinted from Y. Han et al., Phys. Rev. Lett. 104, 017003 (2010). Copyright 2010 by The American Physical Society.]

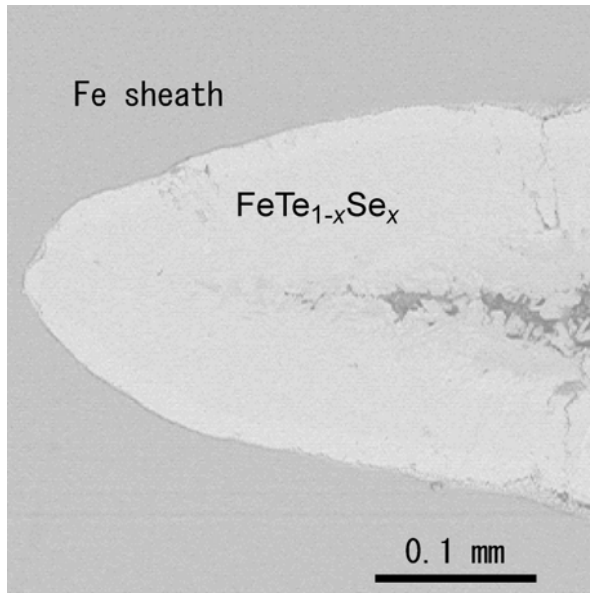

Fig. 53. SEM image of the cross section of the FeTe$_{1-x}$Se$_x$ superconducting wire.

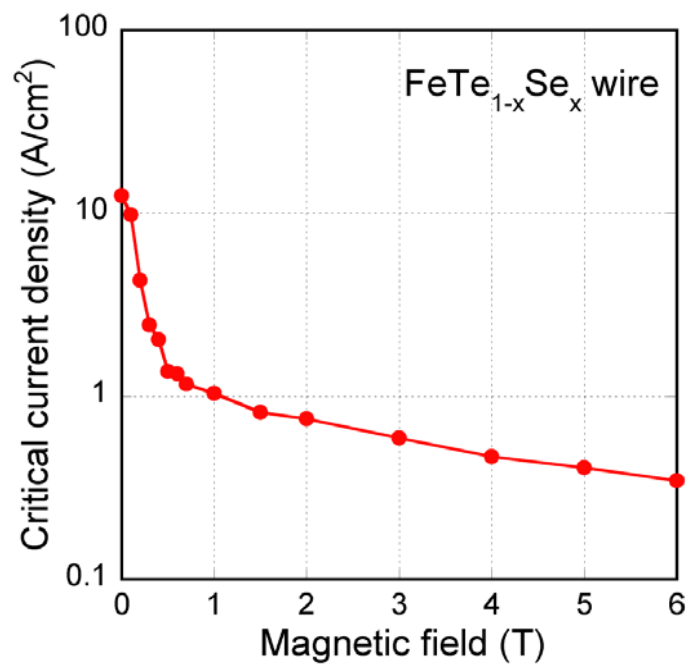

Fig. 54. Magnetic field dependence of critical current density for the FeTe$_{1-x}$Se$_x$ wire.

Acknowledgements

This work was partly supported by Grant-in-Aid for Scientific Research (KAKENHI). The authors thank group members (National Institute for Materials Science) Mr. K. Deguchi, Mr. S. Ogawara, Mr. T. Watanabe, Dr. S. Tsuda and Dr. T. Yamaguchi for useful discussions and experimental help. We also thank the collaborative researchers Dr. H. Kotegawa (Kobe University), Dr. H. Tou (Kobe University), Dr. T. Yokoya (Okayama University), Dr. K. Prassides (University of Edinburgh) and Dr. H. Kumakura (National Institute for Materials Science).